%% file: cms_hot_jupiters.tex
\documentclass[twocolumn]{aastex631}

\usepackage{amsmath}
\usepackage{graphicx}
\usepackage{booktabs}

\newcommand{\tint}{$T_\mathrm{int}$}
\newcommand{\teq}{$T_\mathrm{eq}$}

\graphicspath{{./}{figures/}}

\begin{document}
\title{Tidal Response and Shape of Hot Jupiters}

\author[0000-0003-2451-7939]{Sean M. Wahl}
\affiliation{Department of Earth and Planetary Science, University of California, Berkeley, CA 94720, USA}

\author[0000-0002-5113-8558]{Daniel Thorngren}
\affil{Department of Physics, University of California, Santa Cruz}
\affil{Institute for Research on Exoplanets, Universit\'e de Montr\'eal, Canada}

\author[0000-0003-0834-8645]{Tiger Lu}
\affiliation{Astronomy Department, California Institute of Technology, Pasadena, California 91125, USA.}

\author[0000-0002-7092-5629]{Burkhard Militzer}
\affiliation{Department of Earth and Planetary Science, University of California, Berkeley, CA 94720, USA}
\affiliation{Department of Astronomy, University of California, Berkeley, CA 94720, USA}

\begin{abstract}
We study the response of hot Jupiters to a static tidal perturbation using the Concentric MacLaurin Spheroid (CMS) method. For strongly irradiated planets, we first performed radiative transfer calculations to relate the planet's equilibrium temperature, \teq, to its interior entropy. We then determined the gravity harmonics, shape, moment of inertia, and the static Love numbers for a range of two-layer interior models that assume a rocky core plus a homogeneous and isentropic envelope composed of hydrogen, helium, and heavier elements. We identify general trends and then study HAT-P-13b, the WASP planets 4b, 12b, 18b, 103b, and 121b, as well as Kepler-75b and CoRot-3b. 
We compute the Love numbers, $k_{nm}$, and transit radius correction, $\Delta R$, which we compare with predictions in the literature. 
We find that the Love number, $k_{22}$, of tidally locked giant planets cannot exceed the value 0.6, and that the high \teq~consistent with strongly irradiated hot Jupiters tend 
to further lower $k_{22}$. While most tidally locked planets are well described by a linear-regime response of $k_{22} = 3 J_2/q_0$ (where $q_0$ is the rotation parameter of the gravitational potential), for extreme cases such as WASP-12b, WASP-103b and WASP-121b, nonlinear effects can account for over $10\%$ of the predicted $k_{22}$. $k_{22}$ values larger than 0.6, as they have been reported for planets WASP-4b and HAT-P13B, cannot result from a static tidal response without extremely rapid rotation, and thus are inconsistent with their expected tidally-locked state.


\end{abstract}

\newcommand\RR{{\bf R}}
\newcommand\rr{{\bf r}}

\section{Introduction} \label{sec:intro}

In this paper, we study the response of rotating giant exoplanets to tidal perturbations using the nonperturbative Concentric MacLaurin Spheroid (CMS) method~\citep{Hubbard2013}. The shape of a fluid planet results from a balance of tidal interactions with other celestial bodies and the planets' rotation with self-gravity from the planets' interior mass distribution. While the gas giant planets in our solar system, Jupiter and Saturn, have  had their interiors probed by precise spacecraft gravity measurements \citep{Folkner2017a,Iess2019}, the tremendous distance to exoplanets necessitates more indirect means for studying their interiors. A class of exoplanets that is well suited for the study of tidal interactions is the hot Jupiters, whose relatively large masses and close-in orbits with their host star lead to much stronger tidal interactions than for any planet in our solar system. In addition, intense insolation allows their interiors to maintain much of their primordial heat, with equilibrium temperature, $T_{\rm eq} > 1000$ K \citep{Miller11}, leading to their well-documented, inflated gaseous envelopes \citep{Charbonneau2000,Henry2000,Guillot2002}, which are even more prone to deformation by tidal interactions than colder planets. 

The deformation of a planet can be characterized by the shape of its observable surface, via ratios radii along the of principle axes, $a$, $b$ and $c$, or through normalized moments of its gravity field. For tidal interactions in particular, the first order response is conventionally reported as the second-degree fluid Love number $k_{22}$ \citep{Love1909}.

\citet{Batygin2009} identified a means of constraining $k_{22}$ for HAT-P-13b by considering its special orbital configuration with a highly eccentric outer companion planet. \citet{Buhler2016} applied this technique using observations of secondary eclipses to measure the eccentricity, $e$, and found a $k_{22}=0.31^{+0.08}_{-0.05}$. Meanwhile, \citet{Hardy2017} used independent observations of HAT-P-13b secondary eclipses and inferred a much larger value of $k_{22}=0.81\pm0.10$. \citet{Ragozzine2009} put forward another method for measuring $k_{22}$ by relating it to apsidal precession, which can manifest itself as transit timing variations (TTV) between temporally separated transit observations. 
This method was applied to WASP-4b by \citet{Bouma2019} who estimated a $k_{22}=1.20^{+0.20}_{-0.26}$ from TESS observations showing an offset in transit time with respect to predictions based on observations stretching back to 2007.
Likewise, \citet{Csizmadia2019} estimated $k_{22}=0.62^{+0.55}_{-0.19}$ for WASP-18b using a similar technique, but with apsidal precession rate inferred radial velocity variations (RV) instead. 
An initial suggestion of a detection of apsidal precision for WASP 12-b was ruled out by \citet{Campo2011}.

 The shape of the observable surface of a sufficiently nonspherical planet can be inferred from transit light curves for both fast-rotating oblate planets \citep{Seager2002} and tidally-elongated, prolate planets \citep{Leconte2011,Burton2014}. With the exception of special-case interactions with asecondary companion planet, most close-in hot Jupiters are expected to have evolved to a tidally-locked state with negligible eccentricity \citep{Lin2004,Jackson2008}. Because tidal locking limits the rotation rate to match the orbital period, the tidally induce prolateness is generally more pronounced than the rotational oblateness. For sufficiently close-in hot Jupiters this can lead to a systematic underestimation of reported planetary radius and, by consequence, and overestimation of the bulk density. The effects of rotation and tidal perturbation were analyzed in detail by \citet{Leconte2011} who derived a predictive theory how to correct the observed radii. We will compare our prediction of the radius correction of WASP-12b with their work.  
 \citet{Correia2014} formulated an analytical shape model with an assumed ellipsoidal shape, and a tidal response following the Darwin-Radau equation and calculated $a$, $b$ and $c$ for a number of hot Jupiters, including WASP-4b, WASP-12b, WASP19b and WASP-103b. \citet{Akinsanmi2019} applied the Correia's shape model to a predict the number of transits required to constrain the shape Love number $h_2$ of WASP-103b and WASP-121b for the TESS, PLATO and JWST spacecraft. 
 Similarly, \citet{Hellard2019} and \citet{Hellard2020} predicted the sensitivity of a number of spacecraft, including TESS, PLATO and JWST, to measuring $k_{22}$ for Wasp-121b.

 There is extensive literature on the theory of calculating the shape of a liquid planet, dating back over a century \citep{Jeans1920}, with pioneering calculations on giant planets presented in \cite{Gavrilov1977}. The most commonly used method, known as the theory of figures \citep{zharkov1978}, uses a perturbative approach to determine the planet's response to small deviations of the potential from spherical symmetry. Other works have extended the theory of figures to consider second order effects through higher order perturbative theory \citep{Zharkov2004, Zharkov2010, Correia2013}.  \citet{Padovan2018} adapted a related perturbative method using a matrix-propagator approach, more common in geophysical applications, to exoplanets.
 
 \cite{Hubbard2013} introduced the concentric Maclaurin spheroid (CMS) technique, an nonperturbative, iterative method for more precise calculations of self-consistent shape and gravitational field. The CMS method was subsequently extended to three dimensions and applied to the cases of Jupiter and Saturn \citep{Wahl2016,Wahl2017b,Nettelmann2019,Wahl2020}. In this work we apply the CMS method to hot Jupiter exoplanets for the first time. While more computationally expensive than the theory of figures, the CMS method correctly accounts for non-linear effects that become relevant for extremely deformed planets, most notably effects on the order of the product of rotational  and tidal perturbations \citep{Wahl2017b}. These non-linear effects lead to a splitting of the static $k_{nm}$ with degree $m$, and an enhancement of $k_{22}$ that is significant for Jupiter and Saturn \citep{Lainey2017,Lainey2020,Durante2020,Wahl2020}.

 In addition to the numerical technique, models of shape and gravity also depend on the assumed interior structure and the hydrogen-helium equation of state (EOS). The relationship between tidal response and core mass is discussed in numerous works, \citep[e.g.][]{Batygin2009, Ragozzine2009}, with more centrally concentrated density distributions leading to smaller values of $k_{22}$. Many studies of giant planet interiors employ the semi-empirical \citet{Saumon1995} EOS, while more recent studies have considered equations of state fit to {\it ab-initio} molecular dynamics simulations of hydrogen-helium mixtures based on density functional theory (DFT-MD) \citep{Militzer13, Becker2015, Chabrier2019}, while some theoretical works consider the simpler, more analytically tractable polytropic EOS, \citep[e.g.][]{Leconte2011}. The most pronounced difference between DFT-MD bases equations and \citet{Saumon1995} occurs at pressures of $\sim100$ GPa; as hydrogen transitions from a molecular insulator to an atomic metal \citep{Vorberger2010}, DFT-MD predicts adiabatic temperature profiles that are are cooler and denser.
  \citet{Kramm2012} explored the possible interior structure of giant planet, HAT-P13b using interior models with the \citet{Saumon1995} EOS and the theory of figures to calculate Love number, $k_{22}$.
 \citet{Becker2018} calculated Love numbers using a DFT-MD EOS and the theory of figures for two giant exoplanets Kepler-75b (formerly KOI-889b) and Corot-3b. A similar study of planets in the super-Earth exoplanet regime was carried out by \citet{Kellermann2018}.



\section{Methods} \label{sec:methods}

We first solve the equations of hydrostatic equilibrium for a nonrotating planet as described in~\citet{Seager07}. For the EOS of hydrogen-helium mixtures, we adopt the results from~\citet{Militzer13}, who employed density functional theory molecular dynamics (DFT-MD) simulations to derive an EOS table with absolute entropies at pressure higher than $\sim$5~GPa. At lower pressure, we use the Saumon-Chabrier EOS that was derived with semi-analytical methods~\citep{SC95}. Heavier elements are incorporated into the H-He mixture by following approach in~\citet{hubbard2016}. For the core, we adopted a terrestrial iron-rock ratio of 0.325. For simplicity, we assumed both components are homogeneously mixed. The silicates are described as in \citet{Seager07}. For iron, we employed results from DFT-MD simulations~\citep{WilsonMilitzer2014}. 

For all calculations, we assume a protosolar value for the helium mass fraction, $Y$, of the envelope by setting $Y/(1-Z)=0.27774$~\citep{Lodders10}. The fraction of heavy elements, $Z$, and the entropy, $S$, of the are input
 parameters of our simulations, which govern the density structure of the hydrogen-helium envelope.
  We provide entropy values in units of Boltzmann constant per electron ($k_B/$el), which is referenced to an atomic H:He ratio of 110:9 from~\citet{Militzer13}. In these units, helium rain is predicted to start at $S=7.2$~$k_B/$el, the maximum entropy for which the interior adiabat intersects the pressure-temperature region in Fig.~\ref{fig:PT}, in which hydrogen and helium are predicted to become immiscible~\citet{Morales2010} because hydrogen transitions from an insulating, molecular states to an atomic, metallic fluid~\citep{Vorberger2010} while helium remains in an insulating state. In this work, however, we are primarily concerned with hot Jupiters that we assume to have homogeneously mixed envelopes with entropies $S\ge 7.2$~$k_B/$el.

In addition to the $S$ and $Z$ values for the envelope, adopt values for masses the core and envelope (see Tab.~\ref{parameter}). We integrate the equations of hydrostatic equilibrium starting from a central pressure, $P_c$, to outer pressure boundary, set to 1 bar,  where we assume envelope becomes transparent. We iterate over different $P_c$ values to match the total mass of the planet. In cases for which we have a radius measurement, we iterate over of the core mass or envelope $Z$ to match the planet's mass and radius simultaneously. 

\begin{table*}
\begin{tabular}{m{4cm} m{12cm}}
\hline
Planet mass $M$ [$M_J$] & Input parameter\\
Core mass $M_c$ [$M_\Earth$] & Input parameter. Sets the envelope mass to $M-M_c$. Core shape and radius are derived.\\
\hline
Entropy of envelope $S$ [$k_B/$el.] & Input parameter chosen between 7.2 and 12.0. Sets the temperature-pressure profile of the envelope in Fig.~\ref{fig:PT}. Is derived from $T_{\rm eq}$.\\ 
\hline
Mass fraction of heavy elements in envelope $Z$ & Input parameter that sets also the mass fractions of hydrogen, $X=1-Y-Z$, and helium $Y=0.27774 \times (1-Z)$ because we assume a protosolar helium abundance~\citep{Lodders10}. \\
\hline
Planet radius $a_0$ [R$_J$] & Derived in calculations of nonrotating planet. Alternatively, $Z$ or $M_c$ can be adjusted to match a certain radius.\\
\hline
\end{tabular}
\caption{Parameters of our two-layer interior models.}
\label{parameter}
\end{table*}

The planet mass, $M$, and volumetric average radius, $a_0$, from our calculations of nonrotating planets define the planetary units of mass and length for all following CMS calculations rotating and tidally perturbed planets. In absence of a tidal perturbation, we express the gravitational potential of a axisymmetric rotating planet,
\begin{equation}
  V(r,\mu) =
    \frac{GM}{r} \left[ 1 - \sum_{n=1}^{\infty} \left( \frac{a_0}{r}\right)^{2n} J_{2n}P_{2n}(\mu) \right]\quad,
    \label{eq:potential}
\end{equation}
in terms of the gravity harmonics,
\begin{equation}
  J_n = - \frac{2 \pi}{M a_0^n} \int\limits_{-1}^{+1} d \mu \int\limits_0^{r_{\rm max}(\mu)} \!\!\!\! dr \,\, r^{n+2} \,\, P_n(\mu) \,\, \rho(r,\mu)
  \quad.
\end{equation}
$r_{\rm max}$ defines the outer surface of the planet as function of $\mu=\cos(\theta)$ with $\theta$ being the polar angle. $P_n$ are the Legendre polynomials and $G$ is the gravitational constant. We also define two rotational parameters,
\begin{equation}
    q_{\rm 0} = \frac{\omega^2 a_0^3}{GM} \;\;\; {\rm and} \;\;\; 
    q_{\rm e} = \frac{\omega^2 a_{\rm e}^3}{GM}\;,
    \label{eq:q_0}
\end{equation}
where $\omega$ is the angular frequency of the planet's assumed solid-body rotation. $q_{\rm e}$ is often invoked in the literature when specific planets are discussed for which the equatorial radius, $a_{\rm e}$, is known while $a_0$ is not. For the purposes of this paper, $q_{\rm 0}$ is more convenient because it does not depend on $\omega$ or the equatorial radius, which will only become known once the CMS calculation has converged.

The CMS technique~\citep{Hubbard2013} is a nonperturbative method for deriving the shape and interior structure of rotating planets in hydrostatic equilibrium. Typically one keeps the equatorial radius constant and adjusts the core mass or $Z$ of the envelop to match the observed mass of the planet \citep[e.g.][]{hubbardMilitzer2016}. This approach does not serve our needs because, for different rotation rates, we wish to study what shape is assumed by a planet of given core and envelop masses. The equatorial radii of the planet (and that of the core) are results, not input parameters, of such a calculation. \citet{Hubbard2013} introduced a grid of $\lambda$ points, normalized radii from the planet's center to the equator that anchor the equatorial points of all the equipotential surfaces as the CMS converges towards a self-consistent hydrostatic solution. Here we work with two $\lambda$ grid, one for the core and one for the envelope. Following~\citet{Militzer19}, for the $j$th spheroid surface, $\lambda_j$ is chosen so that a logarithmic grid in density emerges ($\rho(\lambda_j$)/$\rho(\lambda_{j+1})$=const.). We determine the $\rho(\lambda)$ relation from a nonrotating planet calculation, which we then employ for the subsequent calculations with rotation and tides. The density variations throughout core and envelope determine how many grid points must be invested into representing each region accurately. We used 1025 layers in our reported CMS calculations, but already even with 129 layers one obtains good results. For example, for a WASP-12b model with a 0.29 M$_J$ core and 129 layers, we calculate the Love number $k_{22}$=0.16258843 and transit radius correction, $\Delta R$=0.034046. With 1025 layers, we derived $k_{22}$=0.16483250 and $\Delta R$=0.031489, which are both fairly similar. As we will demonstrate, these deviations are small compared to those resulting from changes model assumptions and planet parameters. 

In order to match the core and envelope masses in our CMS calculations of rotating planets, we rescale the $\lambda$ grids of the core and envelope separately as the CMS method converges to a hydrostatic solution. This poses no technical challenges unless the planets are rotating extremely fast ($q_{\rm 0} \gtrsim 0.3)$. 

Once this axisymmetric CMS calculation has converged, we study the planet's shape, compute the gravity harmonics $J_n$, and derive the moment of inertia, $C/(Ma_0^2)$. The hydrostatic structure of axisymmetric CMS calculation also serves as input for a 3D CMS calculation that studies the static tidal response to an external perturber, which could be a the planet's host star, a satellite or a companion planet. 

The CMS technique was extended to three dimensions by \citet{Wahl2017b}. In this version, a third potential term, the gravitational potential from a perturbing mass, $m_S$ at distance, $\RR$, from the planet's center of mass,
\begin{equation}
  W(\rr,\RR) = \frac{Gm_S}{\left|\RR - \rr \right|},
\end{equation} \label{eq:direct_potential}
is added to Eqn.~\ref{eq:potential}, and equipotential surfaces for the combined potential are evaluated on a 3D grid $\rr(r,\mu,\phi)$. \citet{Wahl2020} updated the 3D CMS method by modifying $W$ by subtracting out a linear term determined by an average force,
\begin{equation}
\tilde W(\rr,\RR) = W(\rr,\RR) - \left< {\bf F} \right> \cdot \rr.
\label{eq:av_force}
\end{equation}
This enforces the constraint that the planet's center of mass remains at a specified distance from the perturber. This procedure avoids an issue where the precision of the converged solution is limited by a small shift in center of mass that must be removed each iteration~\citep{Wahl2016,Wahl2017b}.

As with the rotational parameter, we can define two tidal parameters,
\begin{equation}
    q_{\rm tid,0} = - 3 \frac{m_S\, a_0^3}{M\,R^3} \;\;\; {\rm and} \;\;\; 
    q_{\rm tid,e} = - 3 \frac{m_S\, a_{\rm e}^3}{M\,R^3}\;,
    \label{eq:tidal_param}
\end{equation}
where $m_S$ is the mass of the perturber  (the stellar mass for the purposes of this study), and $R$ is the distance to the perturber, in our case the orbital distance. We once again elect to use $q_{\rm tid,0}$ for convenience and refer to simply as $q_{\rm tid}$ for the remainder of the paper. The third and final governing parameter for the tidal calculation is simply the ratio of the planet's radius to the orbital radius, $a_0/R$. In our own solar system, the rapid rotation Jupiter and Saturn place them in the regime where $q_0 \gg |q_{\rm tid}|$, even for their strongest perturbers, Io and Tethys respectively. In contrast, tidally-locked hot Jupiters typically have values of $q_{\rm tid}$ is of a similar or greater magnitude than $q_0$.

As in the axisymmetric case, the converged equipotential surfaces define a density structure from which gravitational field strength can be integrated, now in terms of the tesseral gravity moments $C_{nm}$ and $S_{nm}$ \citep{Wahl2017a}.
For simplicity, we assume that the perturber is in the planet's equatorial plane with $\mu=\cos \theta = 0$. For an exoplanet-star system, this corresponds to a planet with zero obliquity. While this is not a good approximation for the exoplanet population as a whole, it is likely to be the case for many close-in hot Jupiters \citep{Lin2004,Jackson2008}.
We further simplify the geometry by defining the coordinate system such that the perturber is at $\phi=0$, which by symmetry requires $S_{nm}=0$. The tidal Love number can then be expressed as~\citep{Gavrilov1977,zharkov1978},
\begin{equation}
    k_{nm} = -\frac{2}{3}\frac{(n+m)!}{(n-m)!}\frac{C_{nm}}{P_n^m(0)q_{\rm tid,0}}
    \left( \frac{a_0}{R} \right)^{2-n},
\label{eq:kn}
\end{equation}
where $P^m_n(0)$ is the associated Legendre polynomial evaluated at $\mu=0$. For a distant perturber, the m=2 moment dominates the expansion, but higher order moments become more significant as $a_0/R$ increases. We note that the Jupiter and Saturn $k_{22}$ reported later for comparison defined $k_{nm}$ with $q_{\rm tid,e}$ instead of $q_{\rm tid,0}$ \citep{Wahl2017b,Wahl2020}.

In the absence of rotation, $k_{nm}$ is degenerate with respect to
$m$. By contrast, rapidly rotating planets, such as Jupiter an Saturn, are predicted to have have significant splitting of love numbers of the same order $n$ \citep{Wahl2016,Wahl2017b,Wahl2020}. The most readily observable manifestation of this is an enhancement of $k_{22}$ compared to a nonrotating analogue planet, which is evident for observations of both Jupiter \citep{Durante2020} and Saturn \citep{Lainey2017,Lainey2020}.

\section{Results and Discussion} \label{sec:results}

\subsection{General trends}

\begin{figure*}[hbt]
\centering
\includegraphics[width=11cm]{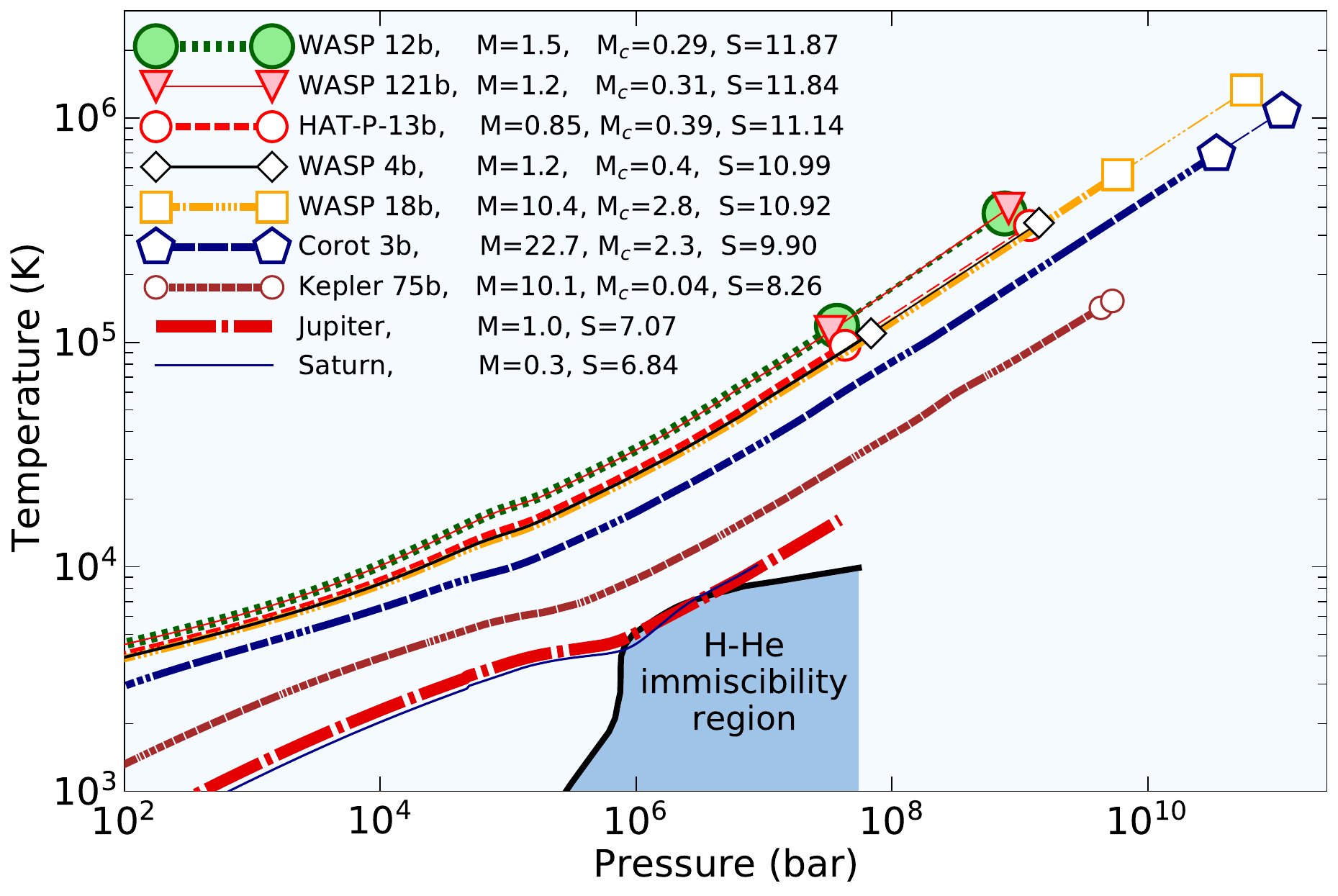}
\caption{ Pressure-temperature conditions in the deep interiors of selected planets. Masses of planets, $M$, and their cores, $M_c$, are given in Jupiter masses in the legend. The two symbols bracket conditions within the planet cores, from the core-envelope boundary (lower left) to the very center (upper right). The distance between the two symbols is primarily controlled by the ratio $M_c/M$. The entropy of the envelope is give in units of $k_b$/el in the legend. Adiabats with $S \le 7.2$~$k_b$/el intersect the hydrogen-helium immiscibility region~\citep{Morales2010}. }
\label{fig:PT}
\end{figure*}

\begin{figure}[hbt]
\centering
\includegraphics[width=0.45\textwidth]{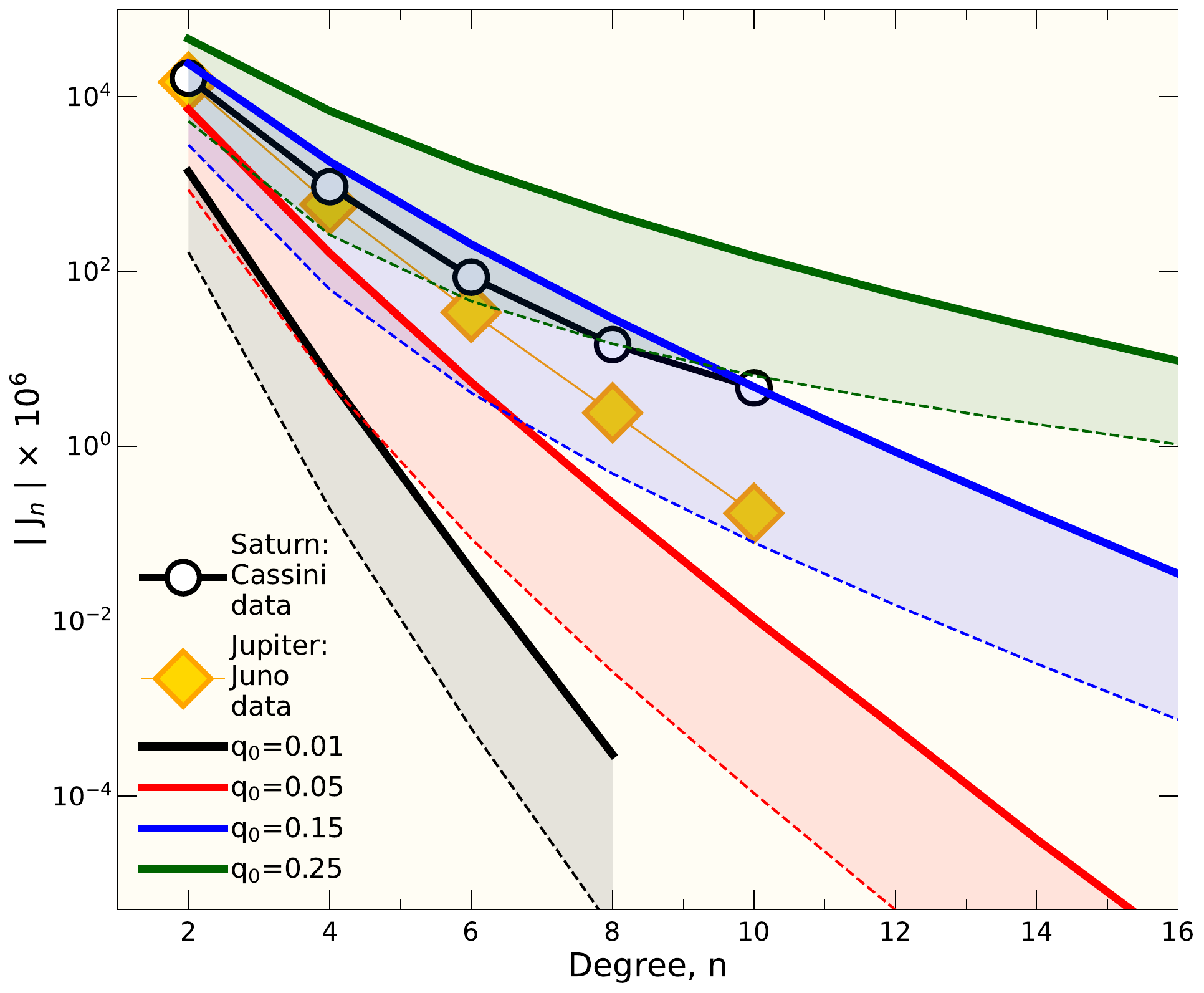}
\caption{ Decay of the gravity harmonics with increasing degree, $n$, is shown for Saturn-mass planets that rotate at different rates. The thick solid lines and thin dashes lines correspond to cold ($S=7.2$~$k_b$/el) and hot ($S=11.0$~$k_b$/el) planets respectively. Spacecraft measurements of Jupiter and Saturn have been included for comparison. The unexpectedly large values of Saturn's $J_6$, $J_8$, and $J_{10}$ have been attributed to differential rotation~\citep{Iess2019}, which we do not include in our exoplanet models because there are no observations to constrain the winds on these planets.}
\label{fig:Jn}
\end{figure}

In Fig.~\ref{fig:Jn}, we compare the gravity harmonics of four Saturn-mass planets with 10 Earth mass ($M_\Earth$) cores as function of the rotational parameter, $q_0$. By definition, all curves start from $J_0=-1$. All cases decay exponentially with increasing degree, $n$, but their decay rates vary with the magnitude of $q_0$. For slowly rotating planets, the $J_n$ decay most rapidly, which is consistent with the fact that all $J_{n\ge 2}$ are zero for a nonrotating planet. With increasing degree, the weight functions of the gravity coefficients become more sharply peaked near the surface~\citep{Militzer2016b}. This means the $J_n$ decay more rapidly for a hot, puffy planet ($S$=11~$k_b$/el) that has less mass near the surface. It also implies that the $J_n$ decay more slowly for fast rotating planets, for which the centrifugal force shifts more mass towards the equator. 

\begin{figure}[hbt]
\centering
\includegraphics[width=0.45\textwidth]{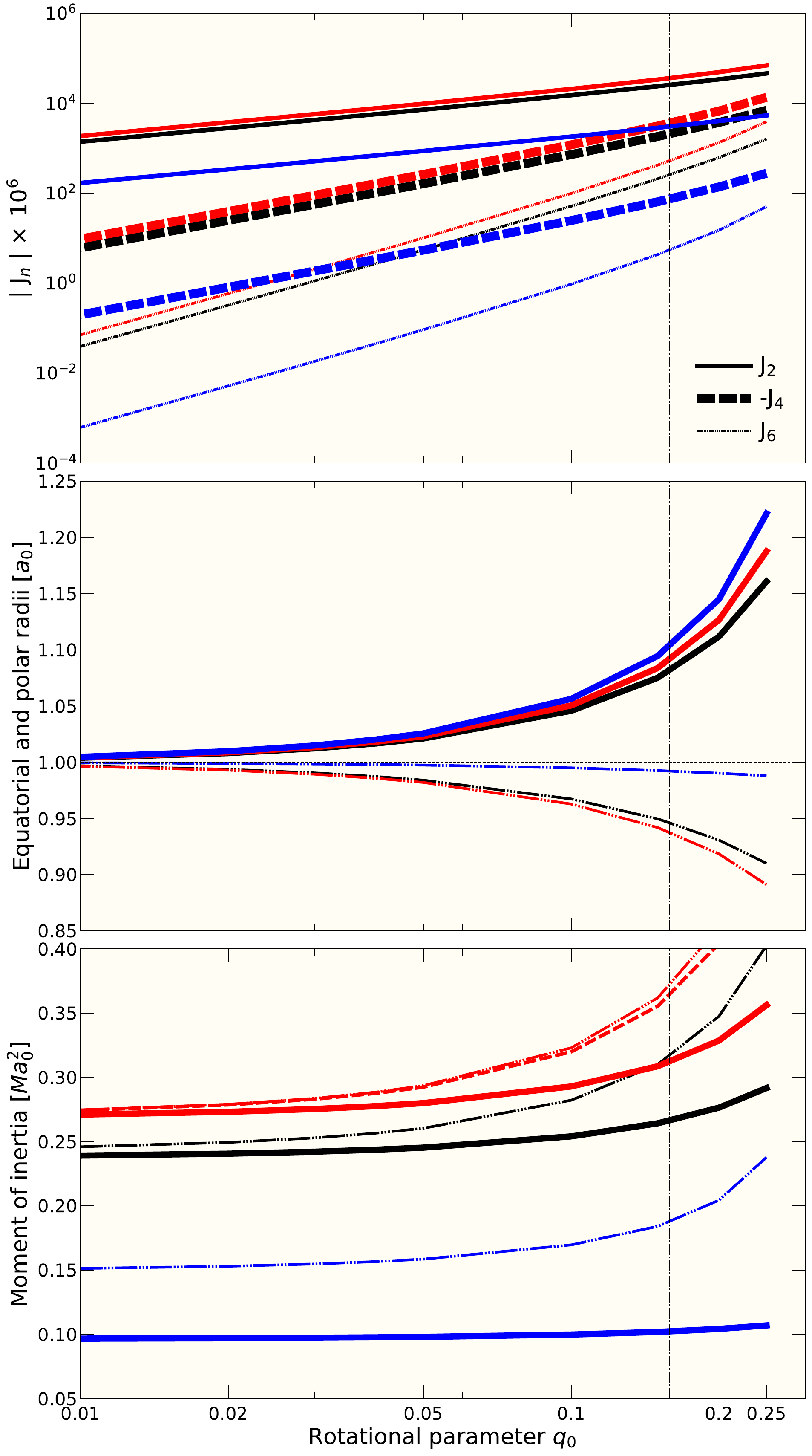}
\caption{Models of three rotating planets without tidal perturbations: cold ($S$=7.2, black, $a_0=0.876\,R_J$) and hot ($S$=11.0~$k_b$/el, red, $a_0=2.604\,R_J$) Saturn-mass planet as well as a cold Jupiter-mass planet (S=7.2~$k_b$/el, blue, $a_0=1.004\,R_J$) are compared as function of $q_{\rm 0}$. The vertical dashed and dash-dotted lines mark the $q_{\rm e}$ values of Jupiter and Saturn respectively. The upper panel shows the gravity harmonics $J_2$, $J_4$, and $J_6$ that scale like $q_{\rm 0}$, $q_{\rm 0}^2$, and $q_{\rm 0}^3$ for small $q_{\rm 0}$ values. The middle panel shows how the equatorial (solid) and polar (dash-dotted lines) radii of these planets change with increasing rotation rate. The radii of the nonrotating planets, $a_0$, are used as normalization. In the lower panel, we compare the CMS predictions (solid lines) for the normalized moment of inertia with the Darwin-Radau expressions: one involving $J_2$ (Eq.~\eqref{DR:J2}, dash-dotted line) and the other relying on the oblateness (Eq.~\eqref{DR:ob}, dotted line).
\label{fig:qRot_plot}}
\end{figure}

In Fig.~\ref{fig:qRot_plot}, we compare various properties of a cold and hot Saturn-mass planet ($S=7.2$ and 11.0~$k_b$/el) as well as a cold Jupiter-mass planet ($S=7.2$~$k_b$/el) each with a 10 $M_\Earth$ core. For moderate values of $q_0$, we find the gravity harmonics, $J_2$, $J_4$, and $J_6$ scale approximately as $q_0$, $q_0^2$, as $q_0^3$, respectively, because with increasing $q_0$, additional mass is shifted towards the equator. For all three planets, we find a sizeable increase in the equatorial radius for large $q_0$. However, the polar radius only shrinks significantly for the two colder planets. The hot Saturn-mass planet is so inflated ($a_0=2.6\,R_J$) and the centrifugal force as large for high $q_0$ that polar radius hardly shrinks as the equatorial radius increases. 

In the bottom panel of Fig.~\ref{fig:qRot_plot}, we plot the moment of inertia. For the hot, puffy Saturn-mass planet, it hardly changes over $q_0$ interval from 0 to 0.25. For the two colder planets, we see a modest increase in the moment of inertia for $q_0 > 0.1$ as the centrifugal force distributes more mass away from the axis of rotation. 

The Darwin-Radau relation gives an approximate expression for the moment of inertia, $C$, of slowly rotating planets ~\citep[see discussion in][]{zharkov1978},
\begin{eqnarray}
    \label{DR:ob}
    \frac{C}{M a_{\rm e}^2} &=& \frac{2}{3} \left( 1 - \frac{2}{5} \sqrt{x} \right)\\
    \nonumber
    x_1 &=& 1+\eta \;\;\;\;\;\; {\rm with} \;\;\;\;\;\;\eta = \frac{5}{2} \frac{q_{\rm e}}{f}\;\;\;\;\;\; {\rm and} \;\;\;\;\;\; f=\frac{a_{\rm e}-c}{a_{\rm e}}\\
    \nonumber
    x_2 &=& \frac{5 q_s}{3J_2+q_s}-1\;\;\;\;\;\; {\rm with} \;\;\;\;\;\; q_s=\frac{\omega^2 s^3}{GM} = q_{\rm e} \frac{s^3}{a_{\rm e}^3}\label{DR:J2}\;.
\end{eqnarray}
where $a_{\rm e}$, $c$, and $s$ are the equatorial, polar, and sphericalized radii. ($\frac{4}{3}\pi s^3$ equals the planet volume.) The quantity $x$ can either be derived from the oblateness, $f$, or expressed in terms of the parameters $J_2$ and $q_s$. Both expressions give similar results unless the density contrast between core and envelope is too large, as we see for the hot Saturn-mass planet in Fig.~\ref{fig:qRot_plot}. In this case, both Darwin-Radau expressions overestimate the moment of inertia by $\sim50\%$ even in the limit of a slowly rotating planet. In this limit, Darwin-Radau results agree fairly well with the CMS predictions of the colder Saturn- and Jupiter-mass planets. The equatorial radii of these two planets are 0.88 and 1.0 $R_J$ while the hot Saturn-mass planet is significantly inflated ($a=2.6\,R_J$), which explains the break-down of the Darwin-Radau expression. Furthermore, one should also be cautious in applying the Darwin-Radau approximation to fast rotating planets like Saturn and Jupiter,  with $q_{\rm e}= 0.158$ and 0.0892 respectively. In this case, $q_{\rm e}$ is no longer a small parameter and the Darwin-Radau assumptions break down.

\begin{figure*}[hbt]
\centering
\includegraphics[width=18cm]{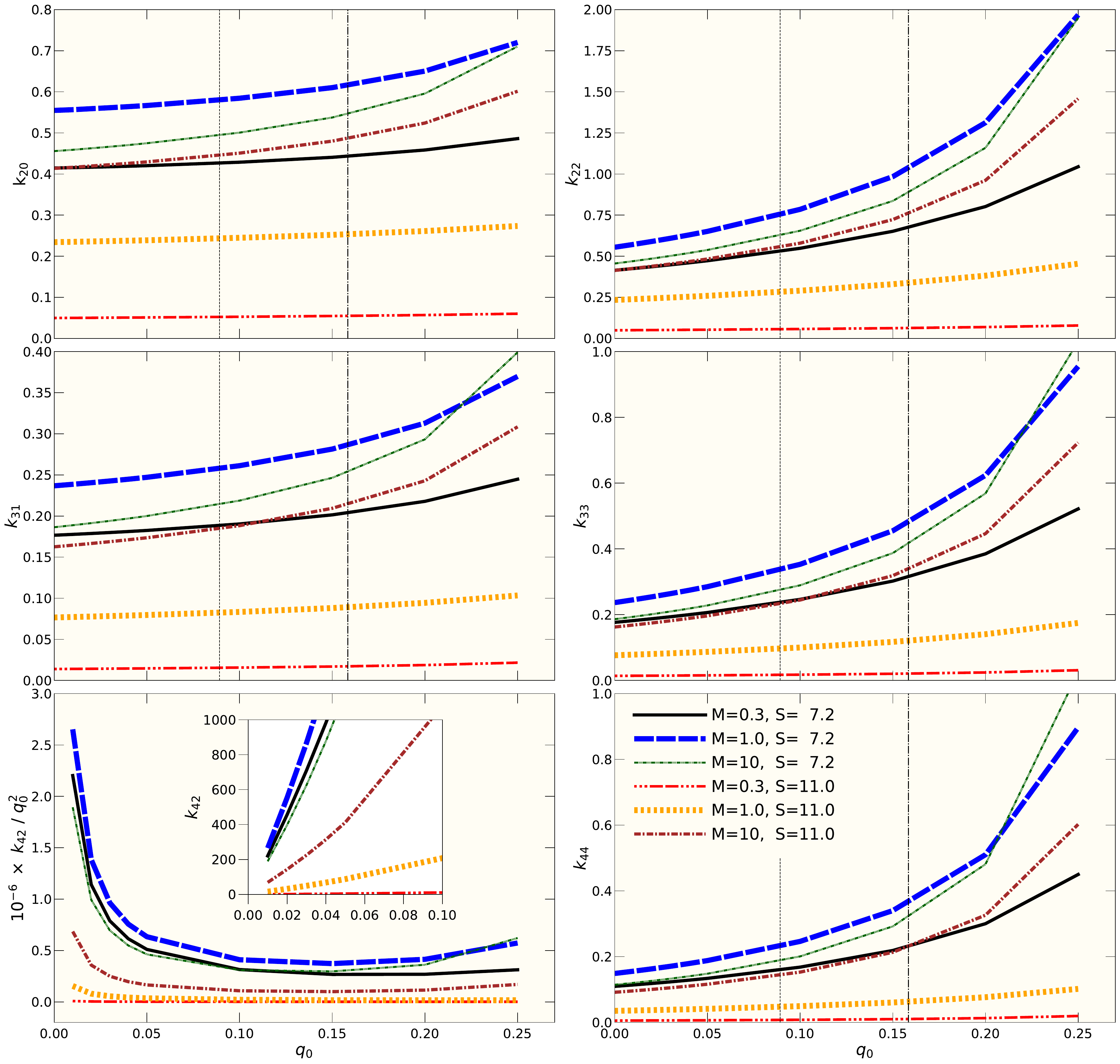}
\caption{Tidal Love numbers, $k_{nm}$ for six planet models are shown as function of $q_{\rm 0}$. The masses and interior entropies are given in the legend of the lower right plot. In the lower left panel, $k_{42}$ has be divided by $q_{\rm 0}^2$ while the inset show the original $k_{42}$. The vertical lines mark the $q_{\rm e}$ values of Jupiter and Saturn as in Fig.~\ref{fig:qRot_plot}. 
\label{fig:knm_plot}}
\end{figure*}

In Fig.~\ref{fig:knm_plot}, we compare the Love numbers, $k_{nm}$, of six planets: three masses 0.3, 1.0, and 10.0 $M_J$ and cold and hot ($S=7.2$ and 11.0~$k_b$/el) interiors. All have 10 $M_\Earth$ cores. In the limit of slow rotation, for given $n$, we find that non-zero $k_{nm}$ values of any $m$ all approach a common value, as expected since angular dependence disappears for a nonrotating. All Love number rise with increasing rotation rate. However, this rise is very small for the hot Saturn-mass planets, which exhibit the smallest Love numbers, followed by the hot Jupiter-mass planet. On the other hand, the cold  Jupiter-mass planet exhibits the largest Love numbers of all six planets, except for very large $q_0$ values where the cold 10 Jupiter-mass planet shows a larger response. For the hot planets (S=11~$k_b$/el), the Love number increases with rising planet mass while no simple trend appears for three colder planets.

\begin{figure}[hbt]
\centering
\includegraphics[width=0.45\textwidth]{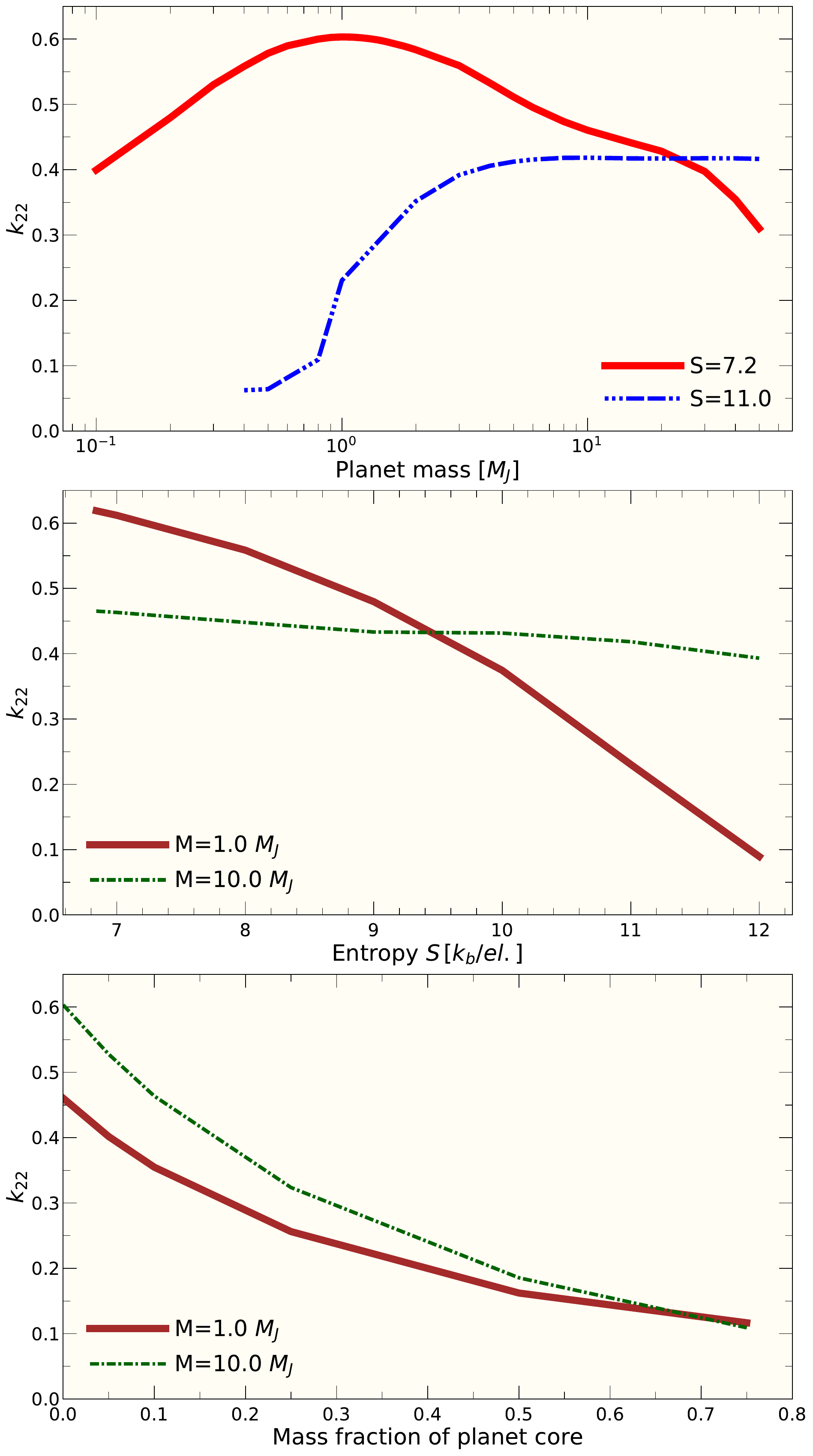}
\caption{ For slowly rotating planets, the three panels show the tidal coefficient $k_{22}$ as a function of planet mass (top), the envelope entropy, $S$ (middle), and  core mass (bottom).} 
\label{fig:k22}
\end{figure}

In Fig.~\ref{fig:k22}, we compare the Love number, $k_{22}$, as function of various parameters in order to motivate $k_{22} \approx 0.6$ as a plausible maximum for slowly rotating planets. In the upper panel, we study the dependence on planet mass. For a cold planets ($S=7.2$~$k_b$/el), $k_{22}$ assumes a maximum value of 0.603 for a one Jupiter-mass planet. This planet has a radius of $1.026\,R_J$, which is close to the maximal radius of $1.069\,R_J$ that emerges for three Jupiter mass planet. 

In the middle panel, we plot $k_{22}$ as function of the envelope entropy, $S$. With increasing $S$, the envelope becomes less dense and thus shows a reduced tidal response. When the entropy of a one Jupiter-mass planet is reduced from $S=7.2$ to a 6.84~$k_b$/el, a typical value for Saturn, $k_{22}$ increases from 0.603 to 0.618. Since this represent cold planet, we argue 0.6 is still a reasonable upper bound for $k_{22}$ of slowly rotating hot Jupiters.

In the lower panel, we study the dependence of $k_{22}$ as a function of core mass fraction while keeping the total planet mass fixed at 1.0 and 10.0 $M_J$.
As expected, $k_{22}$ decreases with increasing core mass fractions because it concentrates more mass in the planet's center where is responds less to tidal perturbations. It is this dependence of $k_{22}$ on core mass that will enable inference of an exoplanet's core mass with future transit measurements~\citep[e.g.][]{Batygin2009, Ragozzine2009}.

\subsection{Results for selected exoplanets}

\begin{figure*}[hbt]
\centering
\includegraphics[width=13cm]{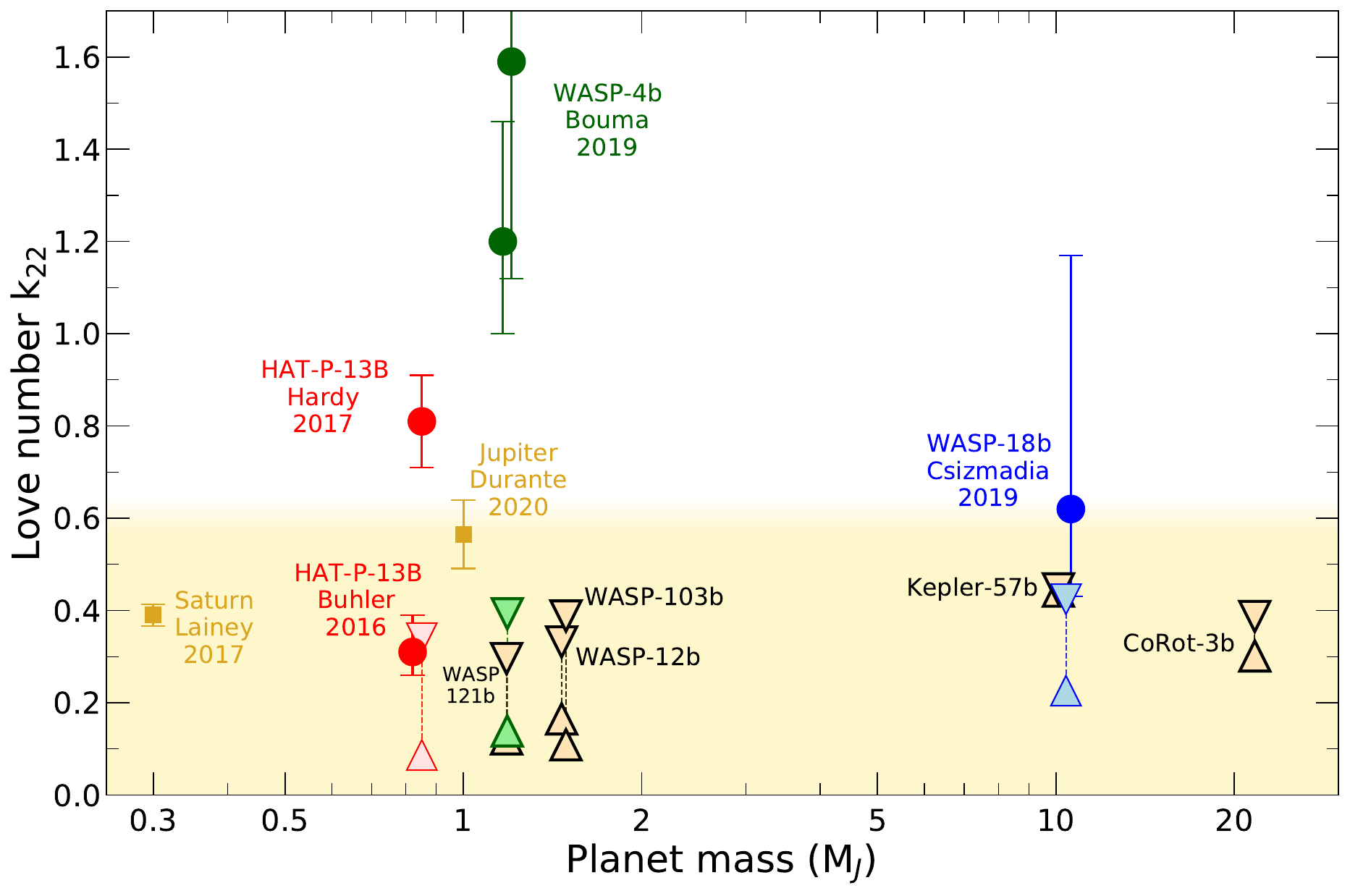}
\caption{ The triangles show the calculated limits on $k_{22}$ for the selected exoplanets in Table~\ref{tab:params}. The upper limit is a model without a central core, while the lower limit is a case with no heavy elements in the envelope ($Z=0$). The filled circles represent exoplanets with reported $k_{22}$ observations. The shaded region represents values of $k_{22} \le 0.6$ that can be explained in terms of a static tidal response.  
} 
\label{fig:comparison}
\end{figure*}

Meaningful models of the interior structure of a given planet require constraints on both planetary mass and radius, which are determined by independent observation techniques. Table \ref{tab:params} summarizes the input parameters for eight exoplanets considered here. Three of the selected planets, HAT-P-13b \citep{Buhler2016,Hardy2017}, WASP-18b \citep{Csizmadia2019} and WASP-4b \citep{Bouma2019} have reported observational constraints on $k_{22}$, while WASP-12b \citep{Campo2011}, WASP-103b \citep{Akinsanmi2019} and WASP-121b \citep{Hellard2020} have each been invoked in studies of the detectibility of $k_{22}$. Figure \ref{fig:comparison} shows reported $k_{22}$ observations compared to the limits on $k_{22}$ we find for all eight selected exoplanets.

\begin{table*}
\input{exoplanet_parameter_table_2021_03_16}

\caption{Parameters for selected exoplanets. (a) \citet{Winn2010}, (b) \citet{Shporer2019}, (c) \citet{Bouma2019}, (d) \citet{Chakrabarty2019}, (e) \citet{Delrez2016}, (f) \citet{Gillon2014}, (g) \citet{Bonomo2015}, (h) \citet{Deleuil2008}. All parameters from \citep{NasaExoplanetArchive}.  $T_{\rm eq,p}$ assumes zero albedo. $S$ interpolated from the relationships in Fig.~\ref{fig:entropy-Teq}.}
\label{tab:params}
\end{table*}

Modelling a realistic interior structure of a giant planet necessarily involves calculating its thermal structure.  For the two-layer models considered, it is natural to parameterize this in terms of the specific entropy of the envelope, $S$.  Cool giant planets like Jupiter and Saturn begin their life with a high specific entropy (from their formation heat) and gradually cool over time \citep[e.g]{Fortney07a}.  Giant planets whose incident flux exceed $2\times10^8$ erg / s / cm$^2$, however, exhibit large radii indicative of hotter interiors than is expected from the physical processes seen in their cooler cousins \citep{Miller11, Demory11}.

\begin{figure}[hbt!]
\centering
\includegraphics[width=0.45\textwidth]{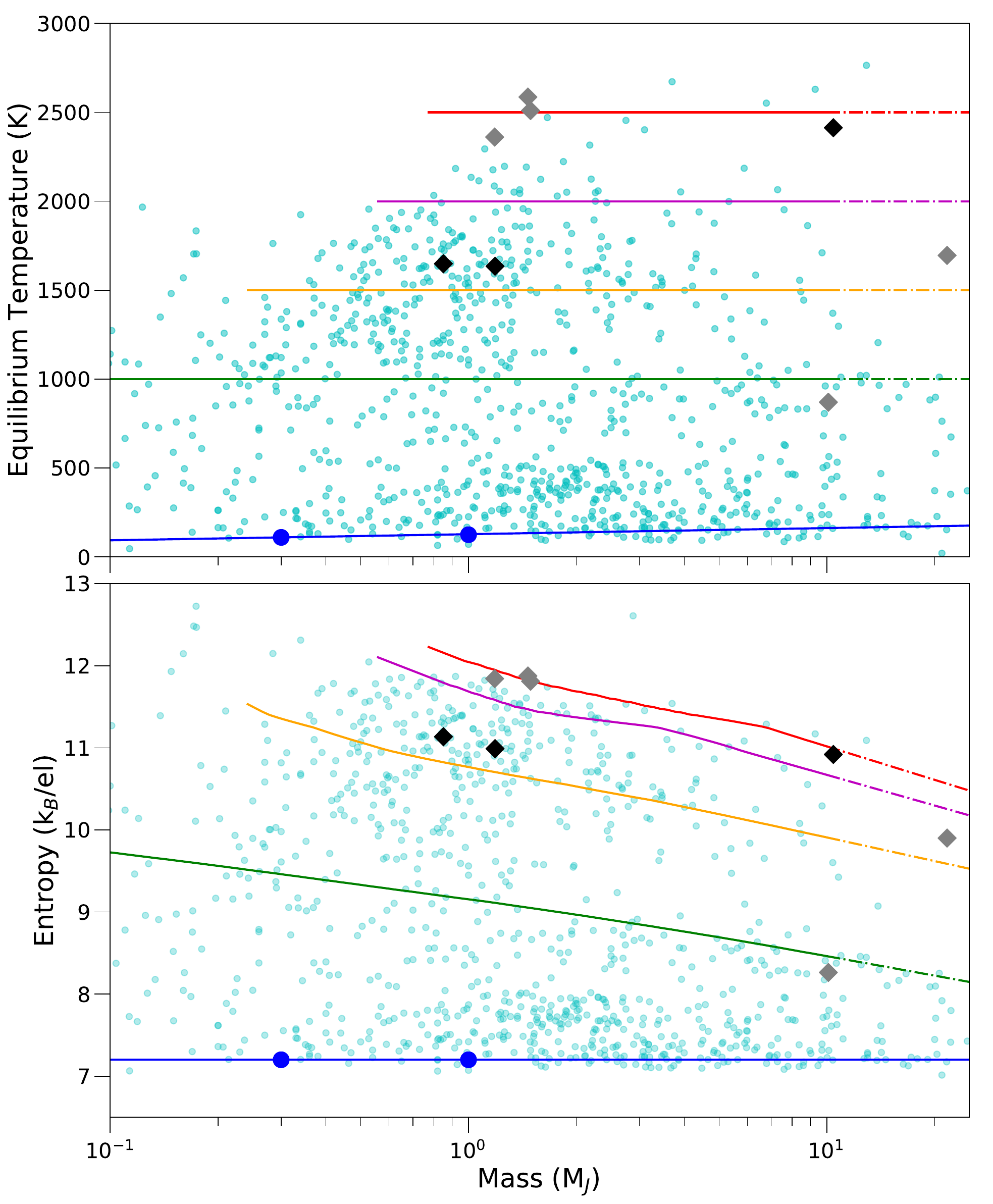}
\caption{ Equilibrium temperature (top) and entropy (bottom) relationship calculated for thermal equilibrium as a function of planet mass shown as solid lines at $T_{\rm eq} = 1000$, 1500, 2000, and 2500 K. Extrapolated \teq-$S$ trends for planet masses $>10$~$M_{\rm J}$ shown with dash-dotted lines. Condition for the onset of helium rain in Saturn and Jupiter denoted by blue circles. Diamonds are selected exoplanets in Tab.~\ref{tab:params} with published $k_2$ observations (black) and without (gray). Planets in \citep{NasaExoplanetArchive} with the necessary parameters are shown in cyan under the same assumptions of thermal equilibrium.
} 
\label{fig:entropy-Teq}
\end{figure}

This hot Jupiter inflation effect may be modeled as an additional heat source within the planet which varies with the incident stellar flux \citep{Thorngren2018, Sarkis2021}.  As they evolve, these planets will, therefore, approach a steady state where the energy flux out of the interior (parameterized by the intrinsic temperature \tint) is equal to the anomalous heating \citep{Thorngren2019}.  Using the atmosphere models of \cite{Fortney2009}, we can relate the envelope's specific entropy to \tint.  Then we apply the results of \cite{Thorngren2018}, which relate the anomalous heating to the incident flux.  The final product is a relationship between the specific entropy of the planet with the incident flux onto the planet (Fig. \ref{fig:entropy-Teq}).  This is helpful because the entropy is extremely difficult to measure observationally, but the incident flux is easily calculated from stellar and orbital properties.  To aide the reader, we represent fluxes as the corresponding equilibrium temperatures assuming zero albedo and full heat redistribution: $T_{\rm eq}^4 = F / (4 \sigma_b)$, where $\sigma_b$ is the Stefan-Boltzmann constant.

Figure \ref{fig:entropy-Teq} shows the relationship between equilibrium temperature and entropy for planets with masses between 0.1 and 10 $M_{J}$ in thermal equilibrium with $T_{eq}$ between 1000 and 2500 K. For planet's within this range the entropy is interpolated with $T_{eq}$ at a constant mass from the two $T_{eq}$-$S$ curves with closest  temperatures. More massive planets were considered by extrapolating the $T_{eq}$-$S$ curves as a function of $\log_{10}(M)$. Cooler planets were considered using a curve constant entropy of $S=7.2$ $k_B/$el., which is the condition for the onset of helium rain as shown in Figure~\ref{fig:PT}. The temperature for this `cold' curve is then defined by the reported effective temperatures for Jupiter and Saturn at the onset of helium rain by \cite{Mankovich20a}. However, planets with lower incoming stellar energy flux are likely to be further from the assumed equilibrium state. 
Our relationship between specific entropy and $T_{eq}$ relies on a fit to the observed population of hot Jupiters \citep[in][]{Thorngren2018}.  As such, we have avoided extrapolating very far into regions of mass-flux space where few to no hot Jupiters are found.  In general, this cutoff moves to higher masses as the temperature increases, as a result of planet formation processes outside the scope of this paper.

Statistical uncertainties in the heating efficiency from \cite{Thorngren2018} are approximately 0.5\%, which leads to an uncertainty in the resulting \tint of around 35 K, varying with the incident flux.  However, these statistical uncertainties are less significant than the modelling uncertainty, which is much more difficult to quantify. An alternate approach was presented by \citet{Sarkis2021} who found broadly similar values for the heating efficiency, though their peak heating of 2.5\% was at a higher temperature of $\sim1860$ K.  \cite{Thorngren2018} found a peak of 2.5\% at $1500-1600$ K. This difference appears to be the result of differences in the atmosphere models. \citep{Molliere2015} included the effects of TiO and VO species on upper atmosphere opacities, whereas \cite{Fortney07a} and therefore \cite{Thorngren2018} did not.  Both papers are fitting to the observed radius via the entropy, so the difference in the predicted entropy for a given planet likely does not differ as much as the heating efficiency. However, this fit relies on assumptions for a planet's composition (abundance of helium and heavier elements) and the equation of states that defines isentropic paths in $P$-$T$ space and the corresponding density. Uncertainties in the equation of state of hydrogen-helium mixtures and their impact on giant planet structure have been discussed by \citet{SG04,Militzer2016b,HelledMazzolaRedmerReview2019}

\begin{figure}[hbt]
\centering
\includegraphics[width=0.45\textwidth]{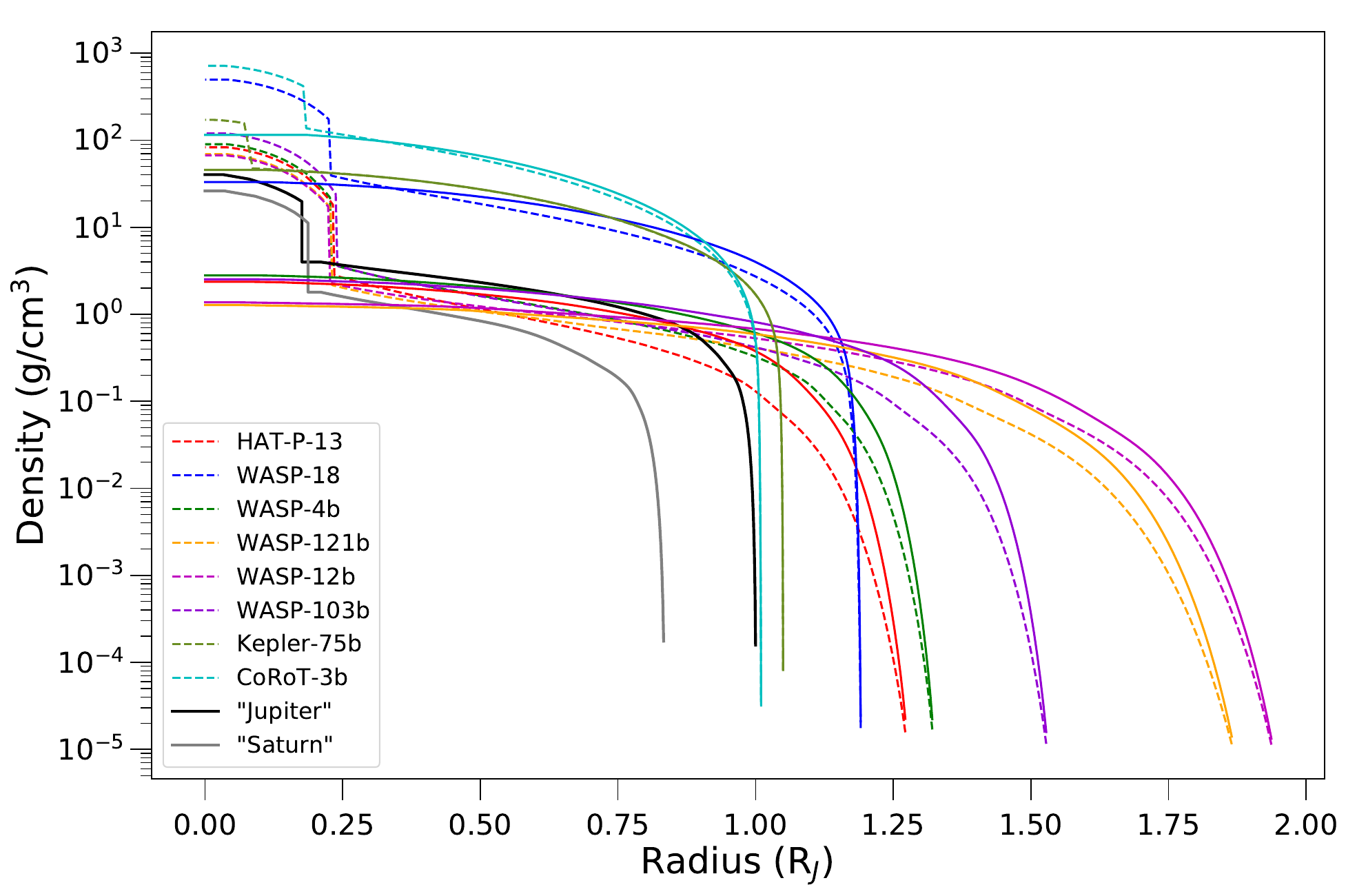}
\caption{ Density as a function of radius from the center of the planet for eight exoplanets and analogue Jupiter and Saturn. For each exoplanet, two end-member interior structures are shown: (solid) fully mixed with no core and maximum envelope $Z$, and (dashed) fully separated with maximum core mass and envelope $Z=0$. Saturn and Jupiter analogues have $M_{c}$ and $Z$ chosen to match the observed $J_2$ \citep{Iess2019,Durante2020}.
} 
\label{fig:profiles}
\end{figure}

The interior density profile is determined by the isentropic pressure-density curve derived from the equation of state, which depends both on the entropy and on the heavy element fraction of the envelope, $Z$. Given the observational constraints on mass and radius, a two-layer model exoplanet can accommodate heavy element mass in both the core and envelope. Figure~\ref{fig:profiles} shows the density profiles of two end-member cases for each selected exoplanet. The first end-member is the case of a core-less model, which corresponds to a maximum value for $Z$ in the envelope. The second case is a model in which $Z=0$ in the envelope, corresponding to a maximum core mass. These are compared to two-layer analogue models of Jupiter and Saturn, which match the observed $J_2$ for each \citep{Iess2019,Durante2020}, in addition to the mass and radius. Two-layer models are known to do a poor job in reproducing the full gravitational field of Jupiter and Saturn, since they ignore the redistribution of helium, as well as a possible `dilute' core \citep{Wahl2017a,Mankovich21} or inhomogeneity of $Z$ across the helium rain layer  \citep{miguel2016,Debras_2019}. For this reason the two-layer analogues of Jupiter and Saturn have more massive central cores and require either negative values of $Z$ or significantly higher temperatures than more complicated interior models. In Figure~\ref{fig:profiles} there are evident influences from both planet mass and equilibrium temperature, with the very massive CoRoT-3b exhibiting the highest densities in the deep envelope, and the highly irradiated WASP-12b and WASP-121b exhibiting far more extended envelopes with notably lower density gradients in the outer portion of the planet.

\begin{table*}
\centering                                                        
\resizebox{\textwidth}{!}{%
\input{exoplanet_results_table_2021_03_30}
}
\caption{Predictioned shape and tidal response for selected exoplanets assuming a tidally-locked state and a two-layer interior models with $N=1025$ CMS layers. For every planet, the top line is for the fully-mixed case with no central core. The lower line corresponds to the fully separated case with envelope $Z=0$ and maximum core mass. 
}
\label{tab:results}
\end{table*}

 As a consequence of their assumed tidally-locked state, both the tidal and rotational parameters, $q_{0}$ and $q_{\rm tid}$, are tied to the orbital distance of the planet. Table~\ref{tab:results} presents $q_{0}$ and $q_{\rm tid}$ for the eight selected exoplanets in the assumed 1:1 resonance locked state, along with results from computed tidal responses, $k_{22}$ and $k_{20}$, and shape for the two end-member interior structures. For the planet shape we report the three principle axes lengths: $a$, equatorial radius along the star-planet axis, $b$, equatorial radius perpendicular star-planet axis, and $c$, polar radius along the rotation axis. 
Additionally, we report the prolateness and oblateness, defined as 
\begin{equation}
    f_{ac} = \frac{a-c}{a} \;\;\; {\rm and} \;\;\; 
    f_{bc} = \frac{b-c}{b}\;,
    \label{eq:oblateness}
\end{equation}
respectively. Although the aforementioned shape parameters are commonly used to describe triaxial ellipsoids, we note that the spheroidal surface predicted by CMS represents a more general shape. In fact, the calculated surface is only an exact ellipsoid in the case of a constant density planet \citep{Wahl2017b}. 


HAT-P-13b and the five planets selected from the WASP catalog
exhibit extremely short orbital periods, and are expected
expected to be tidally locked. Kepler-75b and CoRoT-3b orbit much more distantly, leading to much smaller values of $q_{0}$. For reasonable values of tidal quality factor, $Q$, the two more distantly orbiting plates are also likely to be tidally locked. Alternatively, if like Jupiter and Saturn, these planets rotatation rates have not been significantly slowed by tidal torques, then the values of $q_{0}$ could be orders of magnitude higher than for the tidally-locked state reported here.

For all eight exoplanets, the end-member interior structure case with $Z=0$ in the envelope determines the maximum core mass and radii, and corresponds to a minimum minimum prediction of, $f_{ac}$ and $f_{bc}$, as well as minimal values in $k_{20}$ and $k_{22}$. Conversely, the case with no central core yields the maximum $Z$ in the envelope, as well as a maximum values for $f_{ac}$. $f_{bc}$, $k_{22}$, and $k_{20}$. While not explicitly considered here, planets with dilute cores~\citep{Wahl2017a} are expected to exhibit tidal responses between these two end-member cases, since they represent intermediate degree of concentration of heavy elements between a fully mixed planet, and one with all heavy elements in a dense, central core.

\begin{figure}[hbt]
\centering
\includegraphics[width=0.45\textwidth]{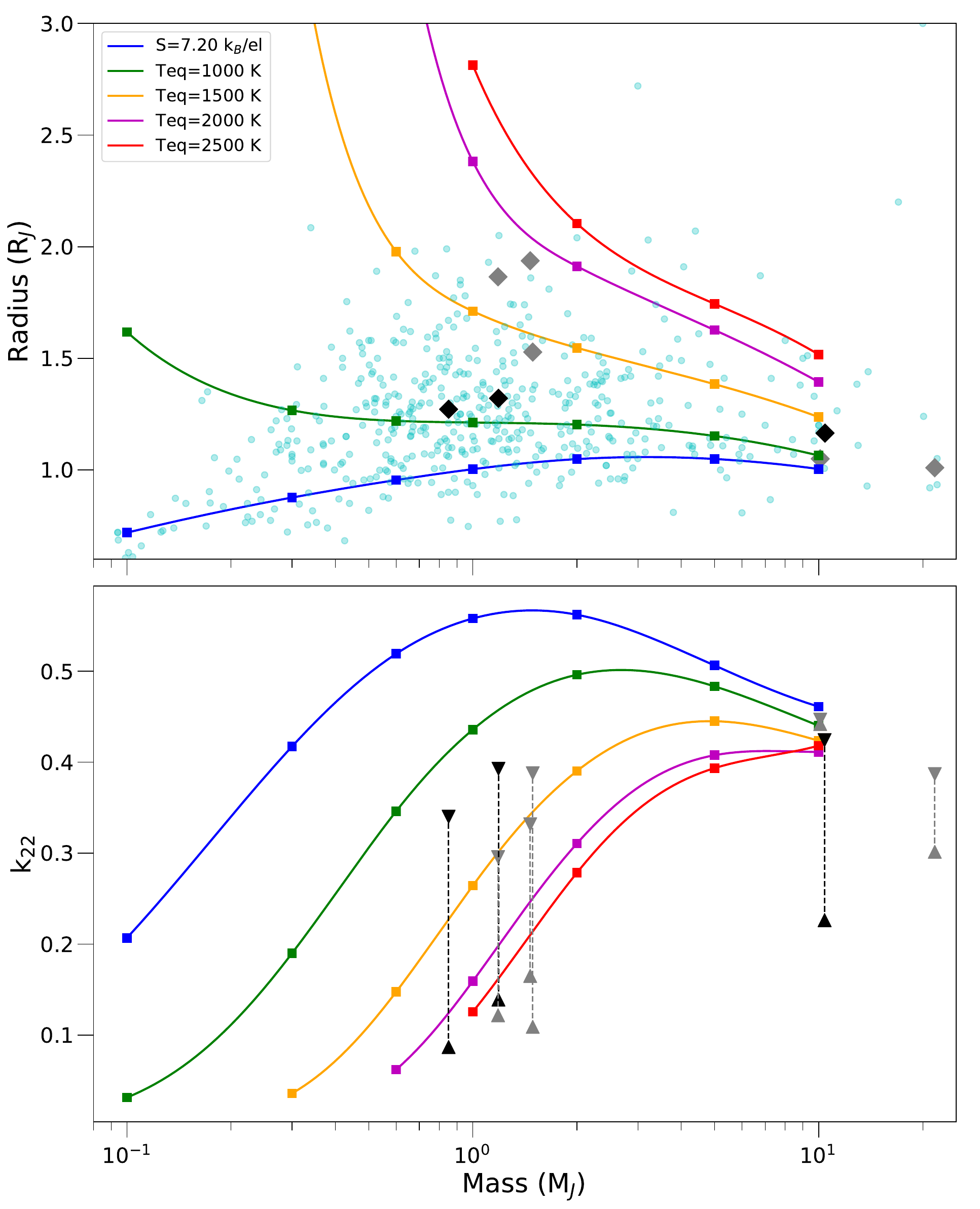}
\caption{ (Top) mass-radius relationship for two-layer interior models with $M_c=10$ $M_\Earth$,$Z=0$ and constant $T_{\rm eq}$ between 1000 and 2500 K (red, magenta, yellow, green squares) and for a constant entropy $S=7.2$ $k_B/$el. (blue squares). Shown for comparison are masses and radii of confirmed exoplanets (cyan circles) \citep{NasaExoplanetArchive} and selected exoplanets (black and gray diamonds). (Bottom) corresponding Love number, $k_{22}$, for mass-radius curves. Predicted range of $k_{22}$ for selected exoplanets shown as in Fig.~\ref{fig:comparison}.
}
\label{fig:mass-radius}
\end{figure}

While the central concentration of heavy elements has the strongest influence on $k_{22}$ for a given planet, the planet mass and equilibrium temperature are significant over the full parameter space of giant exoplanets. Figure~\ref{fig:mass-radius} shows five mass-radius relationships calculated for two-layer models with a constant 10 $M_\Earth$ core and $Z=0$ in the envelope, along with the corresponding $k_{22}$ for a tidal response in the linear regime, $q_{0}<1\times10^{-5}$ and $q_{tid}>-1\times10^{-5}$. The blue curve shows the mass-radius relationship for ``cold" exoplanets with $S=7.2$ $k_{\rm B}/$el. This shows the characteristic trend with radius first increasing with mass to a maximum of $\sim1.05$ $R_{\rm J}$ at $\sim3$ $M_{\rm J}$, and then decreasing as additional mass leads to a shrinking radius due to compaction of the the hydrogen-helium envelope. The calculated $k_{22}$ follows a qualitatively similar trend with a 0.1 $M_{\rm J}$ planet exhibiting a $k_{22}$ of $\sim0.2$ and increasing with mass to a maximum at $\sim$0.56 at $\sim$1.5~$M_{\rm J}$, and then decreasing to $\sim$0.46 at $10$ $M_{\rm J}$. 


The mass-radius relationship for a constant $T_{\rm eq}=1000$ K has a quite different appearance, with an inflated 0.1 $M_{\rm J}$ planet first decreasing with mass, flattening out at $\sim$0.6~$M_{\rm J}$, and then decreasing further for $M>2$ $M_{\rm J}$. The slope at low masses becomes steeper with increasing $T_{\rm eq}$, and the radius decreases monotonically, with an inflection point moving to higher masses for higher \teq. Meanwhile, the calculated $k_{22}$ decreases with increasing \teq, with the maximum $k_{22}$ for a given mass-radius relationship shifting to higher masses. 
While increasing envelope $Z$ leads to a decrease in radius, it has a somewhat less intuitive influence on $k_{22}$. For a colder Jupiter-mass planet with $S=7.20$ $k_B/$el., increasing $Z$ leads to a very slight decrease in $k_{22}$, while an identical mass planet with \teq $=2500$~K exhibits a more substantial increase in $k_{22}$ with $Z$.

The observed $k_{22}$ values for exoplanets, summarized in Figure~\ref{fig:comparison}, cover a wider range of values, most of which are larger than our models can account for with a two-layer interior structure and static tidal response. The \citet{Hardy2017} observation of HAT-P-13b and the \cite{Bouma2019} observation of WASP-4b are both significantly larger than our model predictions, and even considering the reported uncertainties.  They are, in fact, larger than the maximum of $\sim0.6$ for any combination of parameters considered, with the possible exception of models undergoing extremely fast rotation. Our calculated range for WASP-18b has a maximum quite close to lower limit of their prediction with the reported uncertainty \citep{Csizmadia2019}. Given the uncertainties on exoplanet mass, radius and equilibrium temperature be observation may, therefore, be compatible with a static tidal response of a core-less planet, or one in which the core is a small fraction of the planet mass. 

\begin{figure}[hbt]
\centering
\includegraphics[width=0.45\textwidth]{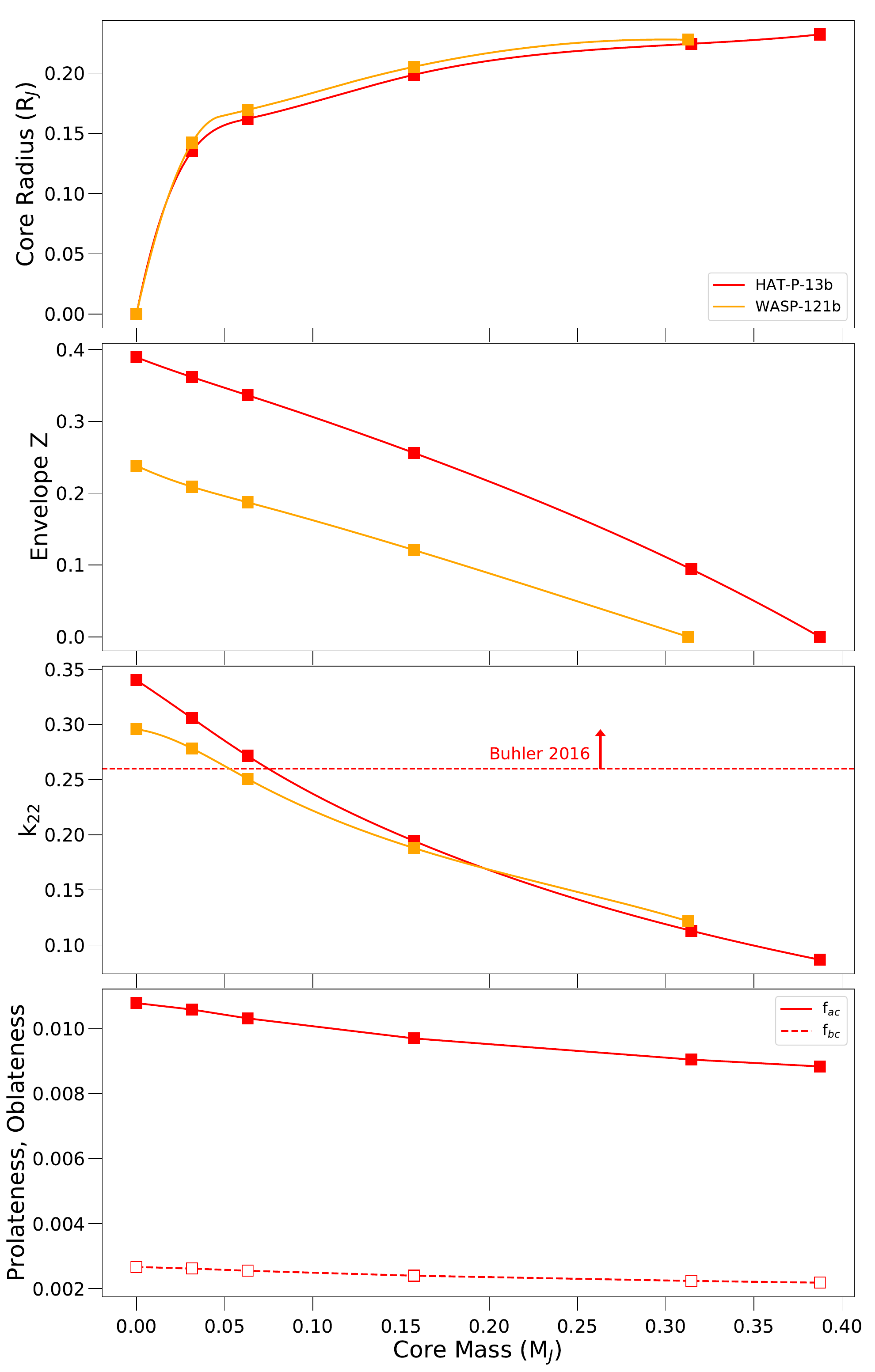}
\caption{ Variation of two-layer model features for HAT-P-13b and WASP-121b with core mass. (Top panel) Core radius, (second panel) heavy element fraction, $Z$, of the envelope, (third panel) Love number, $k{22}$, (bottom) prolateness, $f_{ac}$ (solid) and oblateness, $f_{bc}$ (dashed). The horizontal line in panel 3 shows the minimum value consistent with the observed $k_{22}$ from \citet{Buhler2016}.
} 
\label{fig:compare_mcore}
\end{figure}

The observation of HAT-P-13b by \citep{Buhler2016} is the only observation showing significant overlap with the range of $k_{22}$ predicted here. Figure~\ref{fig:compare_mcore} demonstrates how $Z$, $k_{22}$, and $f$ vary as a function of core mass for our models of HAT-P13b and WASP-121b. The $k_{22}$ observed by \citet{Buhler2016}, is consistent with a model planet ranging from no core to $\sim0.075$ $M_{\rm J}$ ($\sim$24 $M_{\Earth}$) and $Z$ between $\sim0.34-0.39$. It would, therefore, suggest that HAT-P-13b has both a more massive core and an envelope more enriched in heavy elements than Jupiter \citep{miguel2016,Wahl2017a}. The magnitudes of the proplateness and oblateness are governed primarily by $q_0$ and $q_{\rm tid}$, but both shows $~18\%$ decrease between the fully mixed and fully separated HAT-P-13b models. The hotter, more expanded WASP-121b exhibits qualitatively similar dependence on $M_{c}$, but with the observed mass and radius permitting a narrower range of $Z$ and $M_{\rm c}$, and consequently, a smaller range of $k_{22}$.

\begin{figure}[hbt]
\centering
\includegraphics[width=0.45\textwidth]{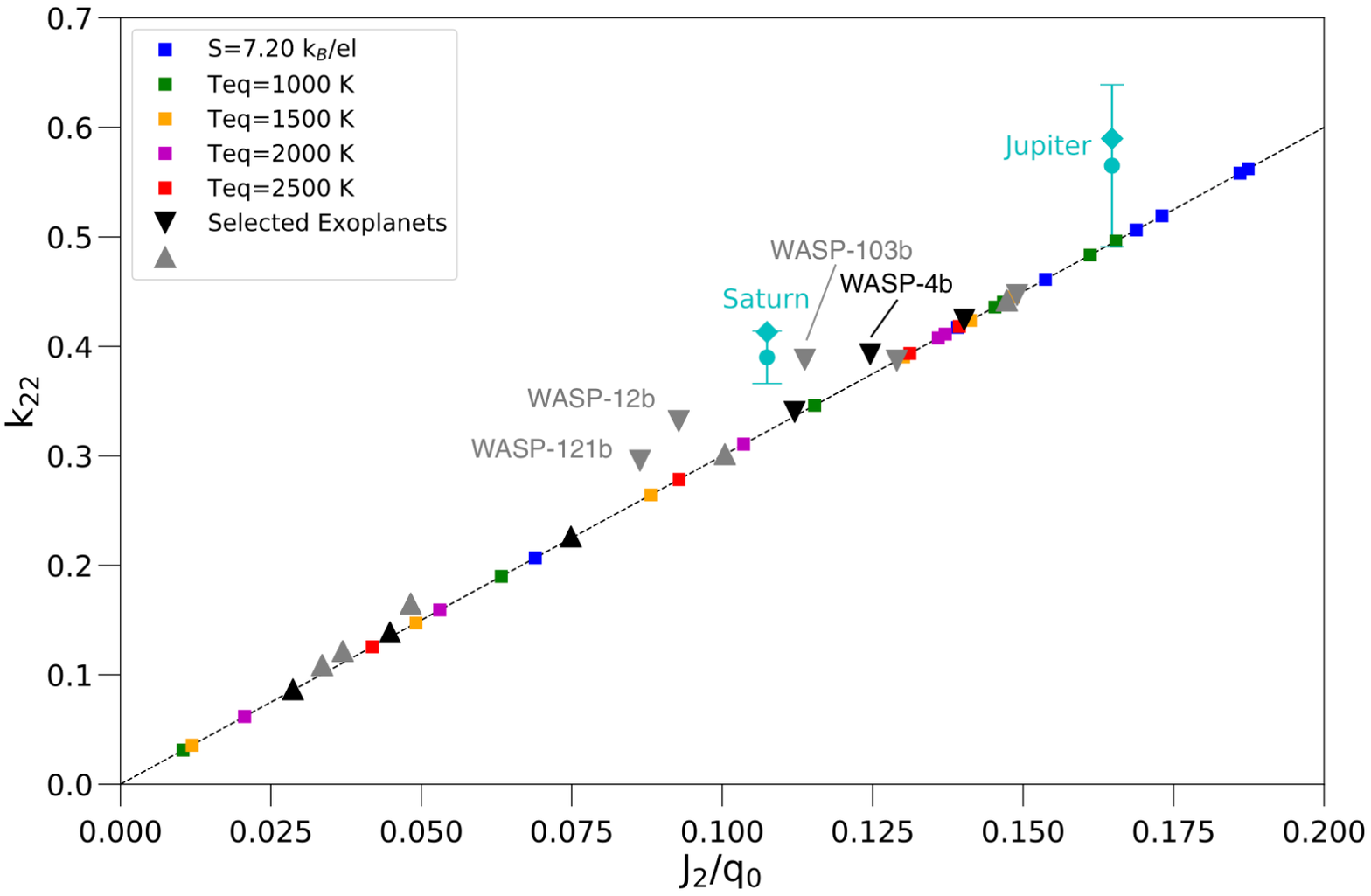}
\caption{ Comparison of the tidal response to the expected linear-regime response. Dotted black line shows the relation $k_{22} = 3 J_2/q_0$. Colored squares show results from the mass-radius curves and, triangles show results for the selected exoplanets following the same notation as Figure~\ref{fig:mass-radius}. Cyan circles and diamond show the observed \citep{Lainey2017,Durante2020} and calculated \citep{Wahl2017b,Wahl2020} static $k_{22}$ for Jupiter and Saturn for comparison. Also labeled are the four exoplanet models with the largest non-linearities.
} 
\label{fig:k22-J2/q}
\end{figure}

In the linear regime, the static tidal response is fully determined by the density profile, and $k_{22}=k_{20}$ \citep{Wahl2017b}. Figure~\ref{fig:k22-J2/q} demonstrates that the simulations for the various mass-radii relationships, calculated with low $q_0$ and $q_{\rm tid}$ precisely follow the relationship $k_{22}=3 J_{2}/q_{0}$. For larger values of $q_{0}$ and $q_{\rm tid}$, non\-linearity in the tidal response leads to splitting between the Love numbers $k_{22}$ and $k_{20}$. Jupiter and Saturn are in a regime with $q_{0} \gg q_{\rm tid}$ and exhibit significant deviation form the linear regime relationship \citep{Lainey2017,Lainey2020,Wahl2017b,Durante2020,Wahl2020}. For hot Jupiters, $q_{0}$ is limited to be of similar order of magnitude as $q_{\rm tid}$ due to tidal locking, which means that significant deviations from the linear relationship occur only in the most extreme cases. Of the selected exoplanets, WASP-12b, WASP-103b, WASP-121b all have $k_{22}$ values enhanced by over $10\%$ from $3 J_{2}/q_{0}$, with WASP-4b a slightly lesser $\sim8\%$ deviation. The largest deviation of $\sim19\%$ is found for a core-less WASP-12 model. Thus we find that in the most extreme cases, tidally-locked hot Jupiters can exhibit appreciably non-linear tidal responses, though still to a lesser extent than the faster rotating solar system giants. Conversely, non-linearities can be safely ignored for exoplanets with $q_{0} \ll 0.01$.

\begin{figure}[hbt]
\centering
\includegraphics[width=0.45\textwidth]{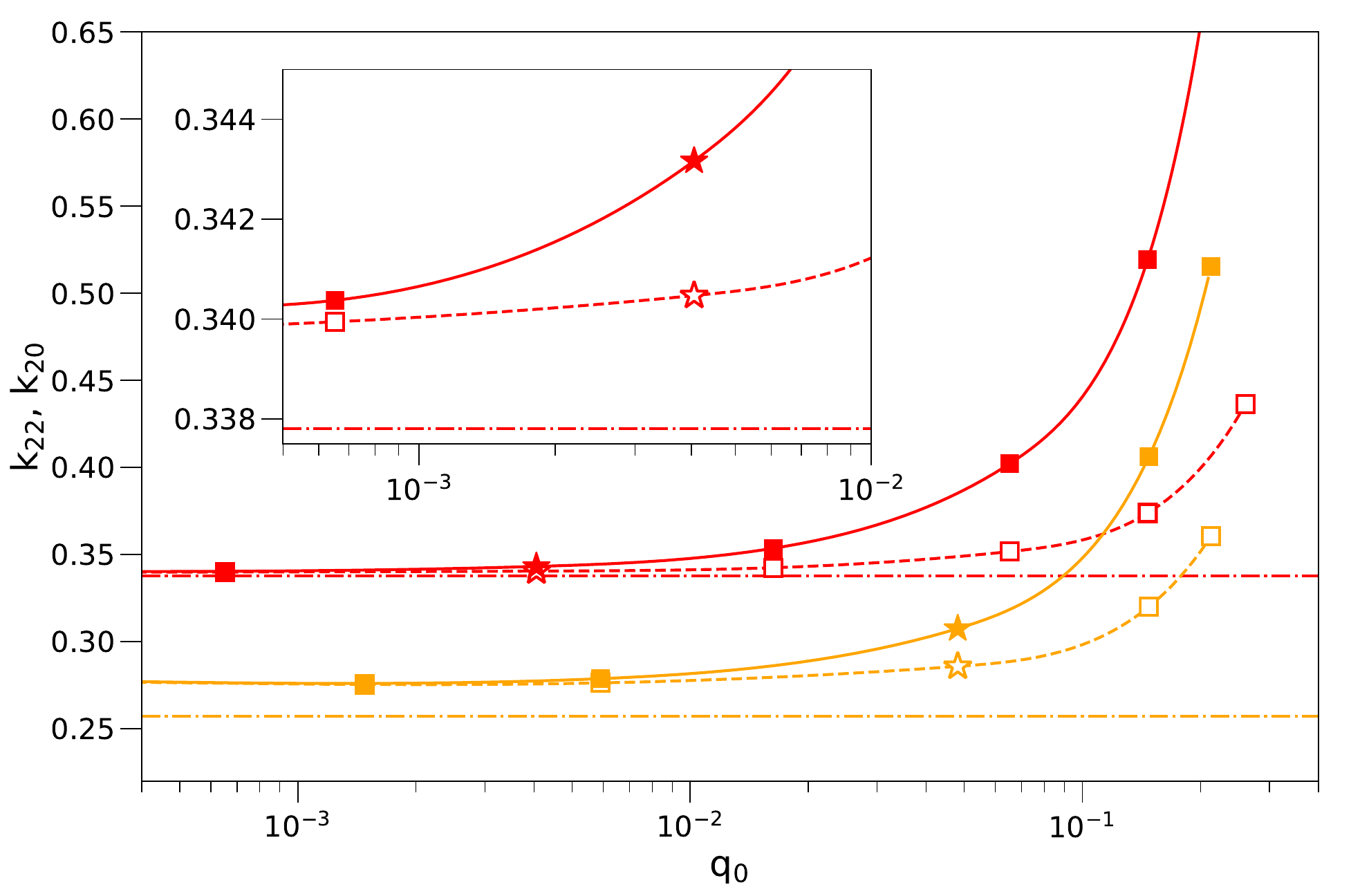}
\caption{ Variation of second-order Love numbers $k_{22}$ and $k_{20}$ with magnitude of rotational parameter $q_0$. Red curves are are for HAT-P-13b, yellow for WASP-121b. $k_{22}$ is a solid curve with filled squares, $k_{20}$ dashed with open squares. $k_{2m}$ for the tidally-locked state with the observed orbital period is denoted with stars. For each planet the linear-regime tidal response of $3 J_{2}/q_0$ is shown with a dash dot. Simulations for this figure use fewer ($N_l=128$) CMS layers.
} 
\label{fig:compare_qrot}
\end{figure}

\begin{figure}[hbt]
\centering
\includegraphics[width=0.45\textwidth]{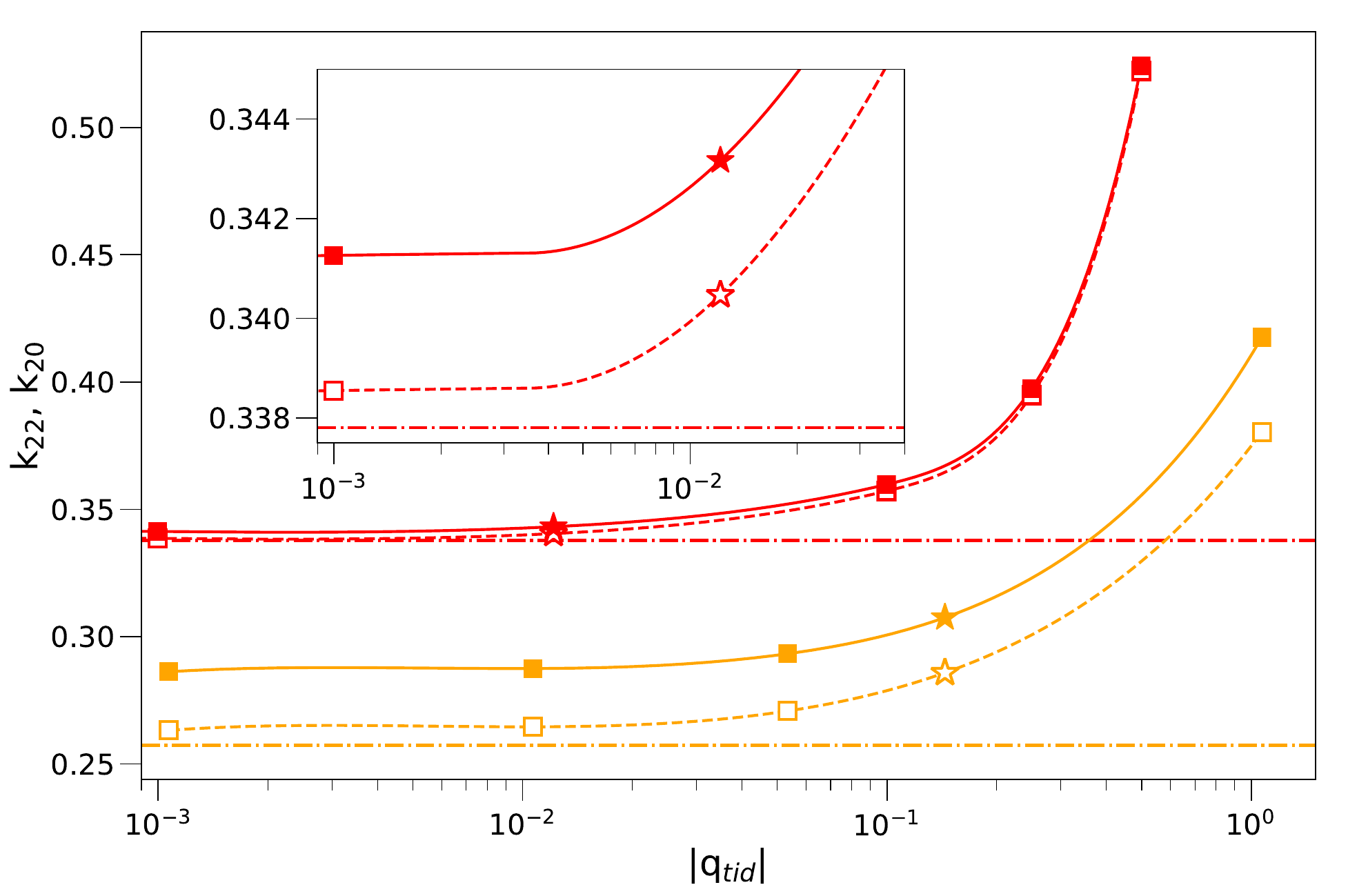}
\caption{Variation of second-order Love numbers $k_{22}$ and $k_{20}$ with magnitude of tidal parameter $q_{\rm tid}$. Quantities are depicted the same as in Figure~\ref{fig:compare_qrot}. Simulations for this figure use fewer ($N_l=128$) CMS layers.
} 
\label{fig:compare_qtid}
\end{figure}

Figure~\ref{fig:compare_qrot} shows influence of $q_0$ on the second-order Love numbers, $k_{22}$ and $k_{20}$, while maintaining $q_{\rm tid}$ at the value in Table~\ref{tab:params} for the HAT-P-13b and WASP-121b models. As $q_0$ is increased,  $k_{22}$ and $k_{20}$ split from their initial degenerate state. While this suggests that an observed $k_{22}$ as large \citet{Hardy2017} might be possible for static tidal response of a planet of HAT-P-13b's interior parameters, it would a require $q_{0}\sim0.2$, roughly two orders of magnitude larger than for the tidally-locked state. Figure~\ref{fig:compare_qtid} considers the complimentary exercise of raising the magnitude of $q_{\rm tid}$ while maintaining $q_0$. In this case the degree of splitting of $k_{22}$ and $k_{20}$ remains similar, while the overall of magnitude increase at large magnitudes of $q_{\rm tid}$. In both cases both $q_0$ and $q_{\rm tid}$ exhibit a comparable effect on the deviation from the linear regime response, but nonlinear response accounting for $\sim$16\% of WASP-121b's $k_{22}$, but $<2\%$ for HAT-P-13b.

When a tidally locked planet transits, the star's flux is dimmed by a disk of radius $\sqrt{bc}$. \citet{Leconte2011} compared this value to the radius of nonrotating, unperturbed planet, $a_0$, and introduced the radius correction factor,
\begin{equation}
    \Delta R = \frac{a_0 - \sqrt{bc}}{\sqrt{bc}} \quad.
\end{equation}
For WASP-12b, the largest correction in their dataset, they obtained a $\Delta R$ range of 0.025--0.035. When we use the same planet parameters, we find reasonable agreement with this correction despite the fact that \citet{Leconte2011} based their work on the theory of figures while we use the nonperturbative CMS theory here. They represented the interiors of hot Jupiters with a polytropic EOS while we used a more realistic EOS for hydrogen-helium-$Z$ mixtures that was derived from {\it ab initio} computer simulations and yielded models for Jupiter and Saturn that agreed well with spacecraft observations~\citep{Iess2018,Wahl2017a}. When we calculated $\Delta R$ for WASP-12b for $M=1.465\,M_J$ and $a_0=1.937\,R_J$, we obtained $\Delta R$=0.0340 for models with $M_c=0.293\,M_J$ and $Z=0$ and $\Delta R$=0.0481 for $M_c=0$ and $Z$=0.19. However, when we adopt the earlier parameters for WASP-12b, $M=1.404\,M_J$ and $a_0=1.736\,R_J$, we respectively obtain $\Delta R$=0.0220 and 0.0330 for our models with and without a core, a difference of only $\sim6\%$ from  \citet{Leconte2011}. In general one finds that giant planets with large cores respond less of tidal perturbations and thus their transit radius correction is smaller.

\citet{Akinsanmi2019} investigated how many transit observations with current instruments would be needed to determine the Love number of WASP-103b. They adopted a very wide range of shape Love number, $h_{22}$, ranging from 0.0 to 2.5 (For fluid planets, $h_{22} = 1 + k_{22}$ holds.) Based on our interior models, we predict a much narrower range for this planet with $k_{22}$ values between 0.109 and 0.388 depending on whether planet has core of $0.588\,M_J$ or no core at all.

\citet{Correia2014} assumed $k_{22}=0.5$ and studied the shape and light curves for a number of exoplanets. \citet{Hellard2019} investigated how $k_{22}$ can be derived from the transit light curves and constructed models for the shape of exoplanets under very specific assumptions. Qualitatively our results for WASP 4b, 12b and 18b in Tab.~\ref{tab:results} agree with these model predictions but one also notices some deviations. \citet{Correia2014} predicted much larger values for the flattenings $f_{ac}$ and $f_{bc}$ of WASP-18b because they used a larger planet radius of 1.52 $R_J$, while we used 1.191 $R_J$. Although we derived a $k_{22}$ range from 0.226 to 0.424 rather than setting it to 0.5. Our predictions for WASP-18b agree fairly well \citet{Hellard2019} with only a small deviation for $f_{ac}$; they derived 0.0077, which is slightly outside of our range of 0.0064--0.0074.

For WASP-4b, the predictions by \citet{Hellard2019} for $f_{ab}$, $f_{bc}$, and $f_{ac}$ are all found to be slightly outside the range that is spanned by our models with and without cores. For example, we determined $f_{ac}$ = 0.036--0.043 while they predicted 0.045. 

For WASP-12b, we find better agreement with the shapes predicted by \citet{Correia2014} once we adopted the older planet radius. However, even in this case their $f_{bc}$=0.031 and that of \citet{Hellard2019}, 0.036, are above of our predicted range of 0.023--0.026 by $\sim20\%$.

\citet{Hellard2020} reported a tentative measurement of WASP-121b Love number, $h_{22}=1.39^{+0.71}_{-0.81}$, which is compatible with the range of $h_{22}-1 = k_{22}$ = 0.122 to 0.296, but not well constrained given the large reported uncertainty.

It is worth noting that in all cases summarized above, once parameters were selected to best match the previous estimates, that our models consistently predict values of $f_{ac}$, $f_{bc}$ and $\Delta R$ slightly below the reported values. This suggests a systematic overestimation of the flattening by these models, possibly resulting from a less realistic hydrogen-helium equation of state, neglecting non-linear effects or artificially or constraining the surface shape to an perfect ellipsoid.

\section{Conclusions}

We studied the tidal response and shape of hot Jupiters, identified a number of general trends, and modeled eight specific exoplanets. Most tidally locked exoplanets are slowly rotating. They are thus in the linear regime, for which Love number is well approximated by $k_{22} = 3 J_2/q_{0}$ and cannot be greater than 0.6. This limit was derived under realistic assumptions for giant planet interiors in which the density strongly varies throughout the envelope. For close-in hot Jupiters, we studies how the high interior temperatures reduce the density profile of the envelope and demonstrate that this change further reduces $k_{22}$. 

We also studied how the tidal response changes with increasing rotation rate. For extremely close-in hot Jupiters, we find that, in spite of tidal locking, rotation rates are sufficient to have a noticeable effect tidal response. For three of the selected exoplanets, WASP-12b, WASP-103b and WASP-121b, we predict $k_{22}$ to exceed the linear value of $k_{22} = 3 J_{2}/q_{0}$ by over 10\%, with the largest deviation of $\sim 19\%$ for WASP-12. 

For realistic planet and stellar parameters, we find $k_{22} < 0.45$ for all eight selected exoplanets.
 This limit is not compatible with much larger $k_{22}$ values that have been reported in the literature of number of hot Jupiters \citep{Hardy2017,Bouma2019,Csizmadia2019}, which may indicate a systematic overestimation of $k_{22}$ by these observation methods. Only the observation of HAT-P-13b by \cite{Buhler2016} overlaps with our predicted $k_{22}$ range. 

If the larger $k_{22}$ observations are confirmed, they imply that these planets are either fast rotating and thus  not tidally locked, or that dynamic tidal effects increase $k_{22}$ in ways that are not understood. For Jupiter, however, dynamic contributions to the tidal response~\citep{Idini2021,Lai2021} have been shown to reduce the static value~\citep{Wahl2020} by approximately $\Delta k_{22} / k_{22}^{\rm static} \approx 4\%$ bringing it in agreement with observations made by the {\it Juno} spacecraft~\citep{Durante2020}. For dynamic tidal effects to explain the discrepancy between the large observed values and our model predictions, they would not only have to have the opposite sign as for Jupiter but also be much larger in magnitude. 

We compare our predictions for the shape of selected planets with earlier models by \citet{Leconte2011,Correia2014,Akinsanmi2019,Hellard2019} that relied on perturbative approaches and simpler assumptions for planetary interiors. Although we find reasonably good agreement for the shape and the transit radius correction if we assume the same masses and radii as the other authors, we also suggest these models may have a small but systematic overestimation of the planet's flattening.

\section*{Acknowledgements}

\begin{acknowledgments}
This work was in part supported by the NASA mission {\it Juno}. BM acknowledges support from the Center for Matter at Atomic Pressure (CMAP) that is funded by the U.S. National Science Foundation (PHY-2020249). DT acknowledges support by the Trottier Fellowship from the Exoplanet Research Institute (iREx). This research has made use of the NASA Exoplanet Archive and data provided by the WASP consortium. 
\end{acknowledgments}

\bibliography{references}
\end{document}

%% file: exoplanet_parameter_table_2021_03_16.tex
\begin{tabular}{llrrrrrrrrr}
\toprule
Name &  $t_{\rm orbit}$ &    $R_{\rm orbit}$  &  $e$  &  $M_{\rm p}$ &  $R_{\rm p}$ &  $M_{\rm st}$  &  $R_{\rm st}$ &  $T_{\rm eff,st}$&  $T_{\rm eq,p}$ & $S$\\
 &  (days) &    (AU) &   &  ($M_{\rm J})$ &  ($R_{\rm J}$) &  ($M_{\Sun}$)  &  ($R_{\Sun}$) &  (K)&  (K) & ($k_B/e^{-}$)\\
\midrule
   HAT-P-13b &       2.916 & 0.04269 &       0.0133 &      0.851 &    1.272 &     1.22 &    1.56 &     5653 &      1649 & 11.14 \\
    WASP-18b &      0.9415 & 0.02009 &       0.0091 &      10.40 &    1.191 &     1.22 &    1.23 &     6400 &      2416 & 10.92 \\
     WASP-4b &       1.338 & 0.02261 &            0 &      1.186 &    1.321 &     0.86 &    0.89 &     5400 &      1634 & 10.99\\
    WASP-12b &        1.091 & 0.02338 &            0 &      1.465 &    1.937 &     1.43 &    1.66 &     6360 &      2585 & 11.87 \\
   WASP-121b &         1.275 & 0.02544 &            0 &      1.183 &    1.865 &     1.35 &    1.46 &     6459 &      2361 & 11.84 \\
      WASP-103b &     0.9255 & 0.01987 &            0 &       1.49 &    1.528 &     1.22 &    1.44 &     6110 &      2509 & 11.81 \\
  Kepler-75b &         8.885 & 0.08164 &         0.57 &       10.1 &     1.05 &     0.91 &    0.89 &     5200 &       870 & 8.26\\
    CoRoT-3b &          4.257 & 0.05738 &            0 &      21.66 &     1.01 &     1.37 &    1.56 &     6740 &      1695 & 9.90 \\
\bottomrule
\end{tabular}

%% file: exoplanet_results_table_2021_03_30.tex
\begin{tabular}{lrrrrrrrrrrrrrr}
\toprule
 Planet &     $q_{\rm rot}$ &      $q_{\rm tid}$ &       $Z$ &  $M_{c}$ &  $R_{c}$ &    $k_{22}$  &    $k_{20}$ &  $a$ & $b$ & $c$ & $f_{ac}$ &    $f_{bc}$ &  $\Delta R$ & $C/(Ma_0^2)$ \\
 name &    &     &        &  [$M_{\rm J}$] &  [$R_{\rm J}$] &     &    &  [$R_{\rm J}$] & [$R_{\rm J}$] & [$R_{\rm J}$] &  &     &   &  \\
\midrule
  HAT-P-13b &  4.06e-2 &   -0.0122 &   0.389 &      0 &      0 &   0.34 &  0.337 & 1.27600 & 1.26561 & 1.26223  &   1.08e-2 &  2.67e-3 &  6.39e-3 &  0.220 \\
            &           &            &       0 &  0.387 &  0.232 & 0.0866 & 0.0862 & 1.27956 & 1.27103 & 1.26825 &  8.84e-3 &  2.18e-3 &  1.86e-3 & 0.108 \\
  \hline
   WASP-18b &  2.61e-2 &  -7.78e-2 &   0.212 &      0 &      0 &  0.424 &  0.422 &  1.19632 & 1.18967 & 1.18748 & 7.39e-3 &  1.84e-3 &  2.04e-3 &  0.240 \\
             &           &            &       0 &   2.53 &  0.225 &  0.226 &  0.225 &  1.19595 & 1.19022 & 1.18832 &  6.38e-3 &  1.59e-3 &  1.45e-3 & 0.177 \\
 \hline
 WASP-4b &   0.0155 &   -0.0464 &   0.308 &      0 &      0 &  0.393 &  0.381 & 1.35600 & 1.31136 & 1.29796 &   4.28e-2 &   1.02e-2 &   1.25e-2 &  0.230 \\
         &           &            &       0 &  0.409 &  0.227 &  0.139 &  0.136 & 1.35369 & 1.31667 & 1.30546 &   3.56e-2 &  8.51e-3 &  7.59e-3 & 0.137 \\
 \hline
 WASP-121b &   0.0481 &    -0.144 &   0.238 &      0 &      0 &  0.296 &  0.275 & 2.02633 & 1.82110 & 1.77207 &    0.125 &   2.69-2 &   3.82e-2 &   0.200 \\
           &           &            &       0 &  0.313 &  0.228 &  0.122 &  0.115 & 2.03212 & 1.84487 & 1.79998 &    0.114 &   2.43e-2 &   2.34e-2 &  0.130 \\
 \hline
 WASP-12b &   0.0594 &    -0.178 &    0.19 &      0 &      0 &  0.332 &  0.304 & 2.16934 & 1.88228 & 1.82032 &    0.161 &   3.29e-2 &   4.64e-2 & 0.206 \\
          &           &            &       0 &  0.294 &  0.223 &  0.165 &  0.154 & 2.17185 & 1.90686 & 1.84932 &    0.149 &   3.02e-2 &   3.15e-2 &  0.150 \\
 \hline
 WASP-103b &   0.0399 &    -0.119 &   0.347 &      0 &      0 &  0.388 &  0.362 & 1.64989 & 1.50378 & 1.46720 &    0.111 &   2.43e-2 &   2.87e-2 & 0.223 \\
          &           &            &       0 &  0.589 &  0.237 &  0.109 &  0.104 & 1.63427 & 1.51422 & 1.48355 &   9.22e-2 &   2.03e-2 &   1.95e-2 & 0.117 \\
 \hline
 Kepler-75b & 2.07e-5 & -6.15e-5 & 4.21e-3 &      0 &      0 &  0.447 &  0.447 & 1.04978 & 1.04973 & 1.04971 & 5.91e-5 & 1.52e-5 & 2.64e-4 & 0.246 \\
           &           &            &       0 &  4.40e-2 & 7.11e-2 &  0.442 &  0.442 & 1.04987 & 1.04982 & 1.04981 &    6.00e-5 & 1.52e-5 & 1.76e-4 & 0.244 \\
 \hline
 CoRoT-3b & 3.74e-5 & -1.11e-4 &  8.32e-2 &      0 &      0 &  0.387 &  0.387 & 1.00894 & 1.00886 & 1.00883 & 1.02e-4 & 2.58e-5 &  1.14e-3 & 0.232 \\
          &           &            &       0 &    2.30 &  0.178 &  0.301 &  0.301 & 1.00992 & 1.00984 & 1.00982 &  9.6e-5 & 2.48e-5 & 1.67e-4 & 0.207 \\
\bottomrule
\end{tabular}